\pdfoutput=1
\def\Ha{H$\rm{\alpha}$}
\def\Hb{H$\rm{\beta}$}
\def\HeI{He\,{\sc i}}
\def\HII{H\,{\sc ii}}
\def\CaII{Ca\,{\sc ii}}
\def\Pg{P$\rm{\gamma}$}
\def\Pb{P$\rm{\beta}$}
\def\FeII{Fe\,{\sc ii}}
\def\NaI{Na\,{\sc i}}

\def\msun{$\rm{M_{\odot}}$}

\documentclass[twocolumn,trackchanges]{aastex63}

\usepackage{longtable}
\usepackage{threeparttable}
\usepackage{etoolbox}
\makeatletter
\patchcmd\@combinedblfloats{\unvbox\@outputbox}{\unvbox\@outputbox}{}{%
  \errmessage{\noexpand\@combinedblfloats could not be patched}%
}%
\makeatother

\newcommand{\chandra}{\textit{Chandra}}
\newcommand{\cxo}{\textit{Chandra} X-ray Observatory}

\newcommand{\err}[2]{\ensuremath{^{_{+#1}}_{^{-#2}}}}
\newcommand{\ee}[2]{\ensuremath{{#1}\!\times\!10^{#2}}}
\newcommand{\Lx}{\ensuremath{L_\mathrm{x}}}

\newcommand{\nH}{\ensuremath{n_\mathrm{H}}}
\newcommand{\pcmsq}{\ensuremath{\mathrm{cm}^{-2}}}
\newcommand{\ergsec}{\ensuremath{\mathrm{erg~s}^{-1}}}
\newcommand{\ergcms}{\ensuremath{\mathrm{erg~cm}^{-2}~\mathrm{s}^{-1}}}

\newcommand{\kms}{\ensuremath{\rm{km~s^{-1}}}}

\graphicspath{{./}{figures/}}

\accepted{April 7, 2020}
\submitjournal{ApJ}

\shorttitle{Early Discovery of SN~2018ivc}
\shortauthors{Bostroem et al.}
\begin{document}

\title{Discovery and Rapid Follow-up Observations of the Unusual Type II SN~2018ivc in NGC~1068}

\correspondingauthor{Azalee Bostroem}
\email{kabostroem@ucdavis.edu}

\newcommand{\UA}{\affiliation{Steward Observatory, University of Arizona, 933 North Cherry Avenue, Tucson, AZ 85721-0065, USA}}

\newcommand{\UCDavis}{\affiliation{Department of Physics, University of California, 1 Shields Avenue, Davis, CA 95616-5270, USA}}

\newcommand{\Granada}{\affiliation{Departamento de Física Teórica y del Cosmos, Universidad de Granada, E-18071 Granada, Spain}}

\newcommand{\IPAC}{\affiliation{Caltech/IPAC, Mailcode 100-22, Pasadena, CA 91125 USA}}

\newcommand{\Trinity}{\affiliation{Department of Physics and Astronomy, Trinity University, San Antonio, Texas}}

\newcommand{\Eureka}{\affiliation{Eureka Scientific, Inc., USA}}

\newcommand{\LCO}{\affil{Las Cumbres Observatory, 6740 Cortona Dr, Suite 102, Goleta, CA 93117-5575, USA}}

\newcommand{\UCSB}{\affil{Department of Physics, University of California, Santa Barbara, CA 93106-9530, USA}}

\newcommand{\UPadova}{\affil{Department of Physics and Astronomy Galileo Galilei, University of Padova, Vicolo dell'Osservatorio, 3, I-35122 Padova, Italy}}

\newcommand{\INAF}{\affil {INAF - Osservatorio Astronomico di Padova, Vicolo dell'Osservatorio 5, I-35122 Padova, Italy}}

\newcommand{\TAM}{\affil{Mitchell Institute for Fundamental Physics and Astronomy, Texas A\&M University,College Station, TX 77843, USA}}

\newcommand{\MIAChile}{\affil {Millennium Institute of Astrophysics (MAS), Nuncio Monse\~nor S\'otero Sanz 100, Providencia, Santiago, Chile}}

\newcommand{\UChile}{\affiliation{Departamento de Astronomía, Universidad de Chile, Casilla 36-D, Santiago, Chile}}

\newcommand{\CTIO}{\affil{Cerro Tololo Inter-American Observatory, National Optical Astronomy Observatory, Casilla 603, La Serena, Chile}}

\newcommand{\ARI}{\affiliation{Aryabhatta Research Institute of observational sciences, Manora Peak, Nainital 263 001, India}}

\newcommand{\Delhi}{\affil {Department of Physics $\&$ Astrophysics, University of Delhi, Delhi-110 007, India}}

\newcommand{\UNC}{\affiliation{Department of Physics and Astronomy, University of North Carolina at Chapel Hill, Chapel Hill, NC 27599, USA}}

\newcommand{\CfA}{\affil{Center for Astrophysics \textbar{} Harvard \& Smithsonian, 60 Garden Street, Cambridge, MA 02138-1516, USA}}

\newcommand{\RaviShankar}{\affil {Pt. Ravi Shankar Shukla University, Raipur 492 010, India}}

\newcommand{\Rutgers}{\affiliation{Department of Physics and Astronomy, Rutgers, the State University of New Jersey, 136 Frelinghuysen Road, Piscataway, NJ 08854, USA}}

\newcommand{\Pulkova}{\affiliation{Central (Pulkovo) Observatory of Russian Academy of Sciences, 196140 Pulkovskoye Ave. 65/1, Saint Petersburg, Russia}}

\newcommand{\IIA}{\affil{Indian Institute of Astrophysics, Koramangala 2nd Block, Bengaluru 560 034, India}}

\newcommand{\Konkoly}{\affiliation{Konkoly Observatory, Research Centre for Astronomy and Earth Sciences, Hungarian Academy of Sciences, H-1121 Budapest, Hungary}}

\newcommand{\UT}{\affiliation{Department of Astronomy, University of Texas at Austin, Austin, TX 78712, USA}}

\newcommand{\Tsinghua}{\affil{Physics Department and Tsinghua Center for Astrophysics (THCA), Tsinghua University, Beijing, 100084, China}}

\newcommand{\JAP}{\affil {Joint Astronomy Programme, Department of Physics, Indian Institute of Science, Bengaluru 560012, India}}

\newcommand{\TAU}{\affiliation{The School of Physics and Astronomy, Tel Aviv University, Tel Aviv 69978, Israel}}

\newcommand{\STScI}{\affil{Space Telescope Science Institute, 3700 San Martin Drive, Baltimore, MD 21218, USA}}

\newcommand{\UCB}{\affil{Department of Astronomy, University of California, Berkeley, CA 94720-3411, USA}}

\newcommand{\Purdue}{\affil{Department of Physics and Astronomy, Purdue University, 525 Northwestern Ave., West Lafayette, IN. 47907, USA}}

\newcommand{\MMT}{\affil{MMT Observatory, P.O. Box 210065, University of Arizona, Tucson, AZ 85721, USA}}

\newcommand{\LBT}{\affil{Large Binocular Telescope Observatory, 933 North Cherry Avenue, Tucson, AZ 85721, USA}}

\newcommand{\Yunnan}{\affil{Yunnan Observatories, Chinese Academy of Sciences, Kunming 650216, China}}

\newcommand{\NAOC}{\affil{National Astronomical Observatories of China, Beijing,Chinese Academy of Sciences, Beijing, 100012, China}}

\newcommand{\NAOCkeylab}{\affil{Key laboratory of Optical Astronomy, National Astronomical Observatories, Chinese Academy of Sciences, Beijing, 100101, China}}

\newcommand{\NAOCkeylabstruc}{\affil{Key Laboratory for the Structure and Evolution of Celestial Objects, Chinese Academy of Sciences, Kunming 650216, China}}

\newcommand{\UOk}{\affil{Homer L. Dodge Department of Physics and Astronomy, University of Oklahoma, Norman, OK, USA}}
 
 \newcommand{\Miller}{\affil{Miller Senior Fellow, Miller Institute for Basic Research in Science, University of California, Berkeley, CA  94720, USA}}
 
 \newcommand{\cifar}{\affil{CIFAR Azrieli Global Scholars program, CIFAR, Toronto, Canada}}

\author[0000-0002-4924-444X]{K. A. Bostroem}
\UCDavis

\author[0000-0001-8818-0795]{S. Valenti}
\UCDavis

\author[0000-0003-4102-380X]{D. J. Sand}
\UA

\author[0000-0003-0123-0062]{J. E. Andrews}
\UA

\author[0000-0001-9038-9950]{S.~D.~Van Dyk}
\IPAC

\author[0000-0002-1296-6887]{L. Galbany}
\Granada

\author[0000-0003-4897-7833]{D. Pooley}
\Trinity
\Eureka

\author[0000-0002-1546-9763]{R. C. Amaro}
\UA

\author[0000-0001-5510-2424]{N. Smith}
\UA

\author[0000-0002-2898-6532]{S. Yang}
\UCDavis
\UPadova
\INAF

%Alphabetical
\author[0000-0003-3533-7183]{G. C. Anupama}
\IIA

\author[0000-0001-7090-4898]{I. Arcavi}
\TAU
\cifar

\author[0000-0001-5393-1608]{E. Baron}
\UOk

\author[0000-0001-6272-5507]{P. J. Brown}
\TAM

\author{J. Burke}
\UCSB
\LCO

\author{R. Cartier}
\CTIO

\author[0000-0002-1125-9187]{D. Hiramatsu}
\UCSB
\LCO

\author[0000-0001-6191-7160]{R. Dastidar}
\ARI
\Delhi

\author[0000-0002-7566-6080]{J. M. DerKacy}
\UOk

\author{Y. Dong}
\UCDavis

\author{E. Egami}
\UA

\author{S. Ertel}
\LBT
\UA

\author{A. V. Filippenko}
\UCB
\Miller

\author[0000-0003-2238-1572]{O. D. Fox}
\STScI

\author{J. Haislip}
\UNC

\author[0000-0002-0832-2974]{G. Hosseinzadeh}
\CfA

\author[0000-0003-4253-656X]{D. A. Howell}
\UCSB
\LCO

\author{A. Gangopadhyay}
\ARI
\RaviShankar

\author[0000-0001-8738-6011]{S. W. Jha}
\Rutgers

\author{V. Kouprianov}
\UNC
\Pulkova

\author[0000-0001-7225-2475]{B. Kumar}
\IIA

\author{M. Lundquist}
\UA

\author[0000-0002-0763-3885]{D. Milisavljevic}
\Purdue

\author[0000-0001-5807-7893]{C. McCully}
\UCSB
\LCO

\author{P. Milne}
\UA

\author[0000-0003-1637-267X]{K. Misra}
\ARI
\UCDavis

\author[0000-0002-5060-3673]{D. E. Reichart}
\UNC

\author{D. K. Sahu}
\IIA

\author{H. Sai}
\Tsinghua

\author[0000-0003-2091-622X]{A. Singh}
\IIA
\JAP

\author{P. S. Smith}
\UA

\author[0000-0001-8764-7832]{J. Vinko}
\Konkoly
\UT

\author[0000-0002-7334-2357]{X. Wang}
\Tsinghua

\author[0000-0002-7531-603X]{Y. Wang}
\NAOCkeylab

\author[0000-0003-1349-6538]{J. C. Wheeler}
\UT

\author[0000-0002-3452-0560]{G. G. Williams}
\UA
\MMT

\author{S. Wyatt}
\UA

\author{J. Zhang}
\Yunnan
\NAOCkeylabstruc

\author{X. Zhang}
\Tsinghua

%% Mark off the abstract in the ``abstract'' environment. 
%==========================
%README BEFORE EDITING
%---------------------------------------------
% Please start each sentence on a new line
% Please use the oxford comma (I'm looking at you Stefano)
% Please use Type II SNe rather than SNe II (and IIL-like and IIP-like)
%==========================
\begin{abstract}
We present the discovery and high-cadence follow-up observations of SN~2018ivc, an unusual Type II supernova that exploded in NGC~1068 ($D=10.1$ Mpc). 
The light curve of SN~2018ivc declines piecewise-linearly, changing slope frequently, with four clear slope changes in the first 30 days of evolution.
This rapidly changing light curve indicates that interaction between the circumstellar material and ejecta plays a significant role in the evolution.
Circumstellar interaction is further supported by a strong X-ray detection. 
The spectra are rapidly evolving and dominated by hydrogen, helium, and calcium emission lines.
We identify a rare high-velocity emission-line feature blueshifted at $\sim7800$ \kms{} (in \Ha, \Hb, \Pb, \Pg, \HeI, \CaII), which is visible from day 18 until at least day 78 and could be evidence of an asymmetric progenitor or explosion.
From the overall similarity between SN~2018ivc and SN~1996al, the \Ha{} equivalent width of its parent \HII~region, and constraints from pre-explosion archival \textit{Hubble Space Telescope} images, we find that the progenitor of SN~2018ivc could be as massive as 52 \msun{} but is more likely $\rm{<12}$ \msun{}.
SN~2018ivc demonstrates the importance of the early discovery and rapid follow-up observations of nearby supernovae to study the physics and progenitors of these cosmic explosions. 
\end{abstract}

%% Keywords should appear after the \end{abstract} command. 
%% See the online documentation for the full list of available subject
%% keywords and the rules for their use.

\section{Introduction} \label{sec:intro}
Single stars with masses greater than $\rm{\sim8}$ \msun{} are thought to explode as core-collapse supernovae (SNe).
Type II SNe, those with hydrogen in their spectra, are empirically classified by the shape of their light curves.
The light curves of Type IIP SNe show a long plateau, $\sim80$--120 days after explosion, before falling by a few magnitudes over $\sim20$ days and eventually settling on a decline powered by the radioactive decay of $\rm{{}^{56}Ni\rightarrow {}^{56}Co\rightarrow {}^{56}Fe}$. 
On the other hand, the light curves of historical Type IIL SNe (e.g., SN~1979C, SN~1980K) decline linearly (in mag day$^{-1}$) before  transitioning to the radioactive decay phase.
While historically these two classes were separated \citep[e.g.,][]{1979barbon, 1993patat, 1994patat, 2012arcavi, Faran2014_IIP, Faran2014_IIL}, today, with larger samples of light curves, we see that these classes blur together: there are intermediate objects showing a linear decline like that of  Type IIL SNe, and a clear fall onto the radioactive decay tail like that of Type IIP SNe \citep{2014anderson,Valenti2015, 2016galbany}. 
Additionally, there appears to be a smooth continuum of slopes during the hydrogen recombination phase.
Throughout this paper we will refer to this collective class as Type IIP/IIL SNe, and we use the Type IIL-like designation to indicate objects that are similar to the historical Type IIL SNe and Type IIP-like to describe SNe that show a clear plateau.

The continuity of Type IIP/IIL SN observational properties makes sense when one considers the physical mechanism producing the plateau in Type IIP-like SNe.
The progenitors of Type IIP/IIL SNe are massive stars with hydrogen envelopes. 
This hydrogen is ionized by the SN shock and as it cools, it recombines, producing a receding recombination front in the expanding ejecta.
The progenitors of Type IIP-like SNe have large hydrogen envelopes and a recession rate which matches the expansion rate, producing a constant luminosity.
If the progenitor has experienced more mass loss, then the hydrogen envelope is less massive.
In this case, the canonical picture is that the photosphere recedes faster, leading to a linearly declining light curve, the steepness of which depends on the amount of hydrogen in the envelope \citep{1971grassberg, 1989young, 1993blinnikov, 2016moriya}.
We note that although this is the standard explanation, the physical mechanism that produces Type IIL-like SNe is not well understood, and it is also possible that increased $\rm{}^{56}Ni$ mass or interaction with circumstellar material (CSM) could produce a linearly decaying light curve.

Over the course of its lifetime, the material lost via winds from the progenitor star leads to a substantial amount of CSM. 
The configuration of the CSM and its density depend on the time of the mass loss, the rate of the mass loss (which could be steady or episodic), and the symmetry of the mass loss. 
Once the star explodes, the radiation, shock wave, and (at later times) ejecta can interact with the material lost prior to explosion, producing observational signatures such as narrow emission lines, enhanced luminosity, and blue colors \citep{2014smith}. 
When narrow or intermediate-width hydrogen emission lines are observed, the SN is denoted a Type IIn SN \citep[see, e.g.,][for a review of SNe]{1997filippenko}. 
Historically, CSM was identified in SNe either through narrow lines produced by unshocked, photoionized CSM, intermediate-width lines created by shocked CSM, or by light curves that deviated from the typical Type IIP/IIL shape and color. 
Recently, signs of interaction have been identified beyond these traditional diagnostics. 
CSM interaction has been seen in Type IIP/IIL SNe at late phases \citep{Andrews2010,Mauerhan2017}, early CSM interaction has been invoked to explain the early-time light curve of most Type IIP/IIL SNe  \citep{Morozova2017, Morozova2018}, and the presence of CSM material very close to the progenitor has been used to explain early, narrow, high-ionization features (``flash spectroscopy'' lines) in several Type II SNe \citep[e.g.,][]{2006quimby, Galyam14, 2015smith, Khazov2016, Hosseinzadeh18}. 
Each of these signatures of interaction occurs during a specific phase of evolution and could be missed without continuous and frequent observations from explosion through the nebular phase.
In order to fully understand the mass-loss history of their progenitors, it is important to follow Type IIP/IIL SNe as close to explosion as possible, for as long as possible, with the highest cadence available.

\begin{figure*}[ht]
\includegraphics[width=18cm]{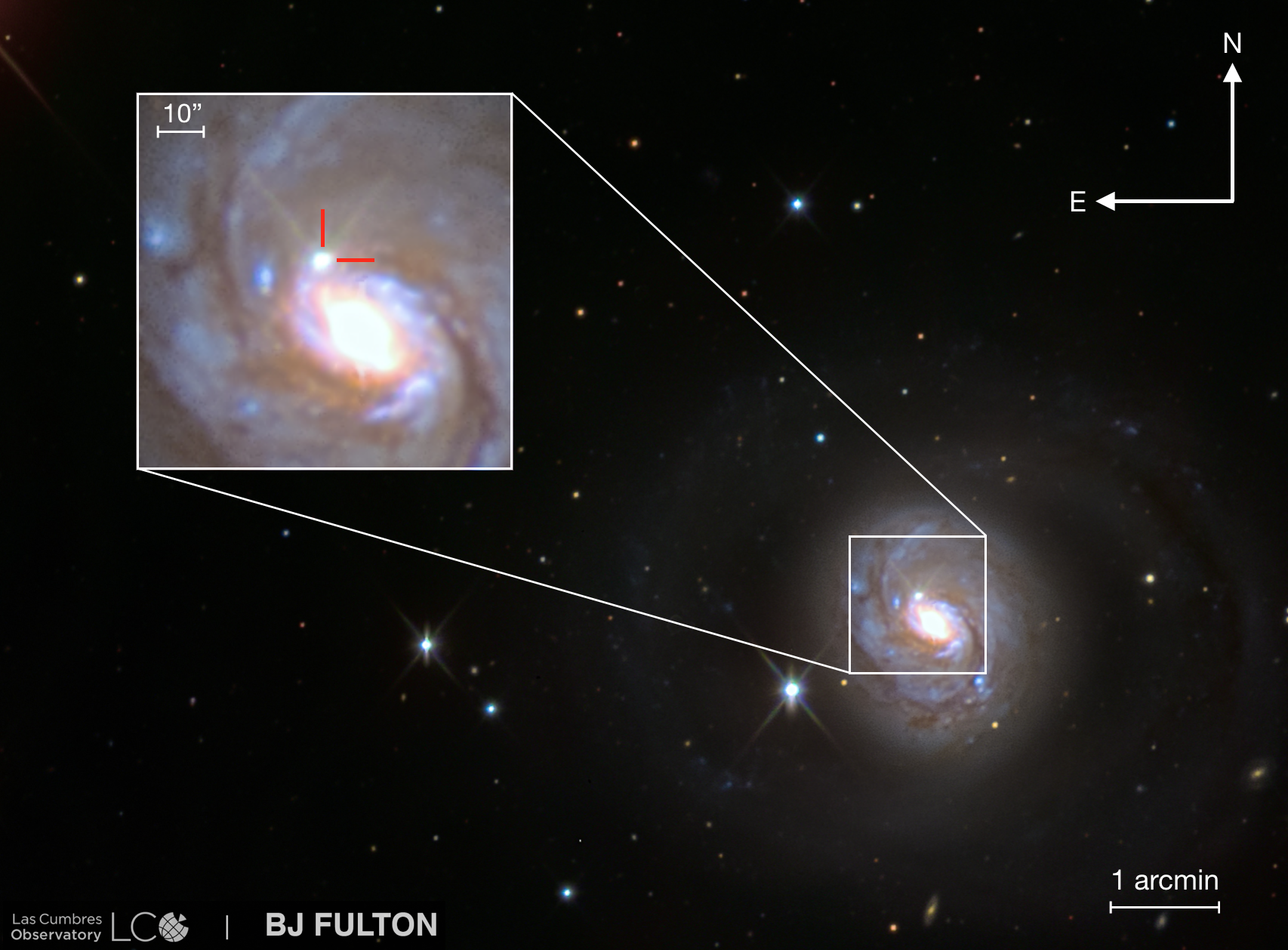}
\caption{A color image of SN~2018ivc and its host galaxy NGC~1068 composed of multiband observations obtained with Las Cumbres Observatory.
The inset shows a close-up image of the nuclear region of NGC~1068.
 SN~2018ivc is indicated with the red guider lines northeast of the nucleus.}
\label{fig:picture}
\end{figure*}

Here, we present observations of SN~2018ivc, a Type IIL-like SN that exploded in a complex CSM environment in the well-studied Seyfert 2 galaxy NGC~1068 (M77; see Figure~\ref{fig:picture}), and was discovered by the $D<40$\,Mpc SN Survey \citep[DLT40;][]{Tartaglia18}. 
We adopt a distance of $D=10.1$ Mpc \citep[$\mu=30.02$ mag;][]{Tully08} and a recessional velocity of 1037 \kms{} ($z$=0.003793; \citealt{1999huchra}) for NGC~1068 in this work.
Our near-daily observations of the SN show the presence of several rapid changes, both in the spectra and in the light curve, which could be easily missed in  undersampled observations of other Type IIL-like SNe. 
In Section \ref{sec:disc} of this paper, we discuss recent improvements to the DLT40 pipeline that enabled the very rapid discovery and follow-up observations of SN~2018ivc.
We then describe the photometric and spectroscopic observations in  Section \ref{sec:obs}, determine the properties of SN~2018ivc and its host galaxy in Section \ref{sec:snparams}, and discuss the spectroscopic and photometric evolution of SN~2018ivc in Section \ref{sec:evolve}.
In Section \ref{sec:analyze}, we present our search for a progenitor in archival images from the \textit{Hubble Space Telescope (HST)}  and our identification of a SN in the literature with a similar evolution.
Properties of the progenitor system are considered in Section \ref{sec:Discussion} and the paper is summarized in Section \ref{sec:conclude}.

%%%%%%%%%%%%%%%%%%%%%%%%%%%%%%%%%%%%
%------------- DLT40 --------------%
%%%%%%%%%%%%%%%%%%%%%%%%%%%%%%%%%%%%
\section{DLT40 Discovery and Rapid Follow-up Campaigns} \label{sec:disc}
\subsection{DLT40 Survey and Recent Improvements}

We briefly summarize the relevant aspects of the high-cadence DLT40 SN survey, the mechanics of which are described in more detail in \citet{Yang17}, \citet{Tartaglia18}, and \citet{Yang19}.
DLT40 is a $\sim12$\,hr cadence search for SNe, targeting galaxies within $D\lesssim40$\,Mpc, designed with the goal of discovering $\sim10$ SNe per year within a day of explosion. 
Since December 2017, DLT40 has operated two nearly identical 0.41~m telescopes, one at Cerro Tololo Inter-American Observatory (CTIO) in Chile and the other at Meckering Observatory in western Australia.  
Each telescope strives to observe the same set of $\sim400$--600 galaxies each night, providing the effective $\sim12$\,hr search cadence.  
The exposure time is 45\,s per field, in a Clear or Open filter, with a typical limiting magnitude of $r \approx 19$ mag in good sky conditions.  
As we describe below, the search for SNe happens in real time, nearly 24 hours a day.

Several recent improvements to the DLT40 pipeline led to the immediate identification of SN~2018ivc as a true SN, and enabled the rapid follow-up campaign.  
First, DLT40 SN candidates are scored in real time using a version of the pixel-based, random forest machine learning algorithm used by the Pan-STARRS1 survey \citep{Wright15}.  
Once a new, strong SN candidate is identified by the DLT40 algorithm, an email alert is immediately sent to the team, and a  follow-up observation with a DLT40 telescope can be urgently requested at the click of a button to verify the new transient.  
If the response to the automated email alert is immediate, the entire process from data taking to automated discovery to confirmation imaging takes approximately five minutes, and truly brings follow-up response into real time.  
Once a new SN has been verified with follow-up imaging, we often trigger a sequence of intranight DLT40 images, which can elucidate rapid light-curve evolution, as was the case for SN~2018ivc.  
Links in the internal DLT40 web pages allow for nearly automated photometric and spectroscopic triggers of new SNe with the Las Cumbres Observatory network of telescopes \citep{Brown_2013}, and we always request a target of opportunity (ToO) sequence of multiband {\it Swift} images to probe the early ultraviolet (UV) light-curve evolution.  
The DLT40 team also announces all verified SN candidates immediately through the Transient Name Service\footnote{https://wis-tns.weizmann.ac.il} (TNS) -- there are no proprietary candidates.

\subsection{SN~2018ivc Discovery and Rapid Follow-up Observations}
The initial DLT40 discovery image of SN~2018ivc (which was given an internal DLT40 designation of DLT18aq) was taken 2018 Nov. 24.04 (UT dates are used throughout this paper) by the PROMPT5 telescope on CTIO, with a magnitude of $r=14.65\pm0.02$.  
The DLT40 machine-learning algorithm identified a strong SN candidate in the initial image of the NGC~1068 field, and an email alert was sent out two minutes later.  
One of us (R. C. A.) quickly inspected the SN candidate on the DLT40 internal web site and triggered a second observation, which was taken 14\,min after the discovery image.  
Once the SN candidate was confirmed, we took a sequence of PROMPT5 images of the field over the next $\sim2.5$\,hr, during which the SN brightened from the initial $r=14.65$ mag to $14.61$ mag.  
During this time period, we reported the SN candidate to the community \citep{Valenti18} and also triggered the Las Cumbres Observatory network of robotic telescopes to obtain  multiband observations ($UBVgri$) as soon as possible (first images on 2018 Nov. 24.11, 2\,hr after discovery). 
We obtained our first spectrum with the FLOYDS spectrograph on 2018 Nov. 24.28 and a second spectrum with the BFOSC spectrograph on 2018 Nov. 24.70 \citep{2018zhang}.

The last nondetection in the NGC~1068 field by the DLT40 team was on 2018 Nov. 19. 
The field of NGC~1068 had not been observed in the days just prior to the discovery of SN~2018ivc owing to the lunar angle constraint DLT40 places on its target fields ($\theta_{\rm Moon}>25^\circ$). 

%%%%%%%%%%%%%%%%%%%%%%%%%%%%%%%%%%%%
%--------- Observations -----------%
%%%%%%%%%%%%%%%%%%%%%%%%%%%%%%%%%%%%
\section{Observations} \label{sec:obs}
The photometric and spectroscopic follow-up observations of SN~2018ivc were obtained and coordinated through Las Cumbres Observatory's Global Supernova Project (GSP; PI: D. A. Howell), a key project to collect densely sampled optical light curves and spectra of nearby and bright SNe \citep[e.g.,][]{Szalai19,Andrews19}. 

In addition to the extensive optical dataset, we also obtained early UV and near-infrared (NIR) photometry and NIR spectroscopy throughout the first $\sim50$ days, and an early X-ray observation\footnote{We also obtained Giant Metrewave Radio Telescope observations.  However, we found issues with the calibration which rendered the observations unusable.}.

%--------- Photometry ------------%
\subsection{Photometry} \label{sec:obs_phot}
Starting within hours of discovery, high-cadence optical photometric data from the Las Cumbres Observatory telescope network were acquired in the $UBVgri$ bands for SN~2018ivc; all data were reduced using the {\tt lcogtsnpipe} software suite \citep{Valenti_2016} on difference images.  
Photometric monitoring with the Las Cumbres Observatory was stopped when we were no longer able to detect the SN against the bright host background, on 2019 Jan. 21.18 (day 60). 
These observations were supplemented with high-cadence photometry from the DLT40 survey, observed with the Open/Clear filters and calibrated to the {\it r} band. 
These observations were reduced using the {\tt DLT40 pipeline} \citep{Tartaglia18} and photometry was performed on difference images.
DLT40 monitoring was stopped when the SN disappeared behind the Sun.

\begin{figure}[ht]
\includegraphics[width=\columnwidth]{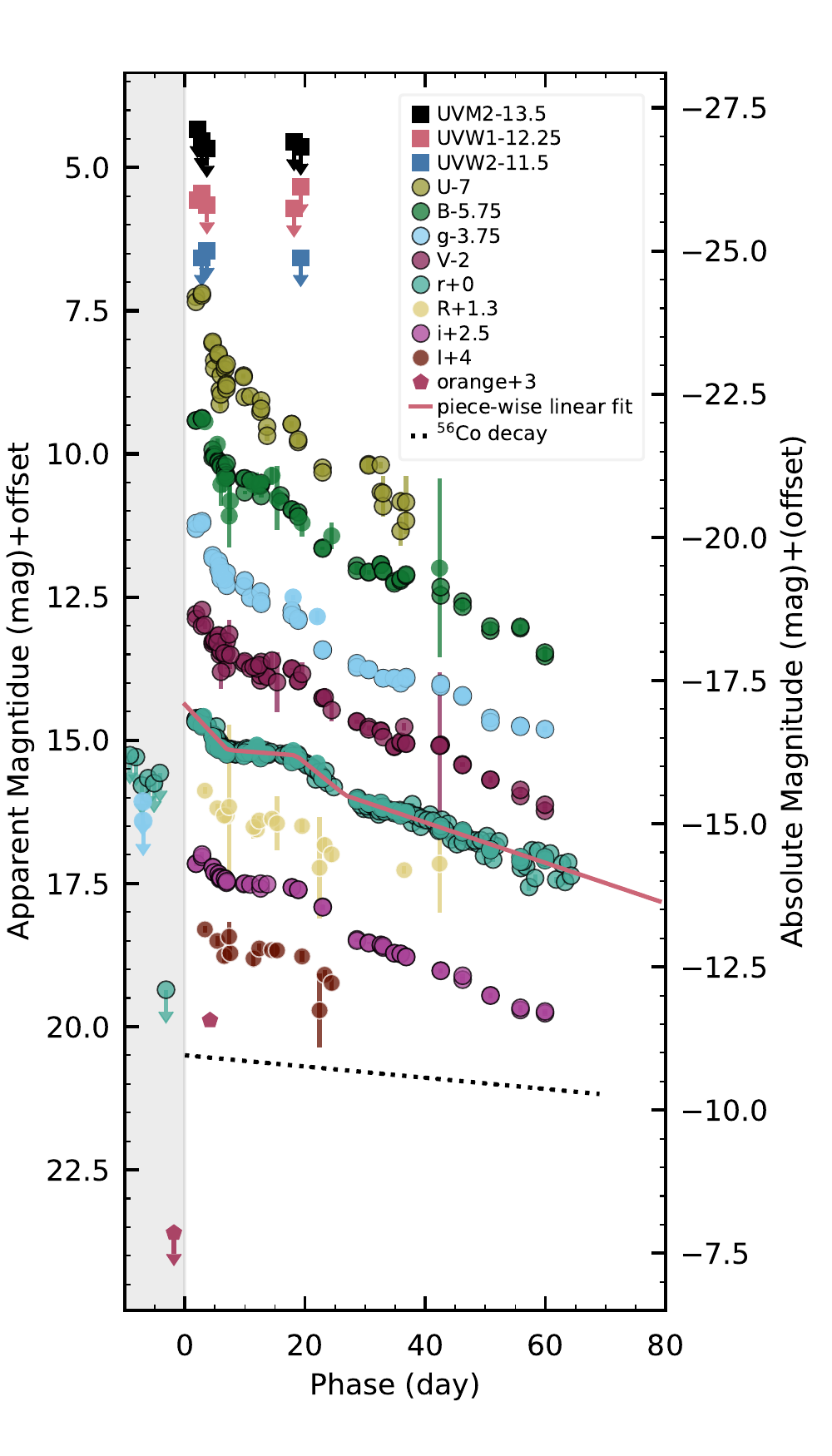}
\caption{The light curve of SN~2018ivc at UV (square symbols) and optical (circles and pentagons) wavelengths.
Difference-image photometry is indicated by a black outline.
Upper limits are denoted with arrows.
All phases are calculated with respect to the inferred explosion epoch (2018 Nov. 22.25; see Section \ref{sec:snparams}) and the period prior to this time shaded in gray.
The piecewise linear fit of the DLT40 \textit{r}-band light curve is shown in pink.
Individual filters have been offset by a constant (denoted in the legend) for ease of viewing. 
The black dashed line shows the V-band slope expected if the light curve is fully powered by the radioactive decay of ${}^{56}$Co. 
}
\label{fig:lc}
\end{figure}

We augment these observations with data from the 1.04~m Sampurnanand Telescope in the \textit{BVRI} bands \citep{1999Sagar}, the 1.3~m Devasthal Fast Optical Telescope in the \textit{BVRI} bands \citep{2012Sagar}, the 2.01~m Himalayan Chandra Telescope (HCT) in the \textit{BVRI} bands \citep{Prabhu_etal_2010}, the Mont4K instrument on the 1.55~m Kuiper Telescope in the \textit{UBV} bands, and the 0.6~m Super-LOTIS telescope \citep{Williams08} in the \textit{BVRI} bands.  
Point-spread-function (PSF) photometry was performed on the original images using the  {\tt DAOPhot} \citep{Stetson1987} {\tt PyRAF}\footnote{{\tt PyRAF} is a product of the Space Telescope Science Institute, which is operated by AURA, Inc., for NASA} package.
As the photometry of these observations was not performed on difference images, imperfect background subtraction produces more scatter in the light curves. 
The optical light curve is presented in Figure \ref{fig:lc}.

Late-time optical observations were obtained with {\sl HST\/} on 2019 July 1 using the Wide Field Camera 3 (WFC3) UVIS channel (F555W and F814W), as part of the ToO program GO-15151 (PI: S.~Van Dyk).
We used {\tt Dolphot} \citep{Dolphin2000, Dolphin2016} to extract the photometry from the FLC frames.
A list of all photometric observations is available in the electronic version of Table \ref{tab:phot}.

SN~2018ivc was also observed with the Neil Gehrels \textit{Swift} Observatory \citep{Gehrels2004}. 
Observations with the Ultra-Violet Optical Telescope (UVOT; \citealp{Roming05}) were reduced and analyzed using the pipeline for the \textit{Swift} Optical Ultraviolet Supernova Archive (SOUSA; \citealp{Brown_etal_2014_SOUSA}), which includes an arithmetic subtraction of the underlying host-galaxy flux measured from pre-explosion imaging.  
The magnitudes use the updated calibration from \citet{Breeveld10} and are on the UVOT/Vega system.  
Optical magnitudes are not reported because the underlying host galaxy was too bright to correct for the coincidence loss.  
Upper limits in \textit{uvw2} and \textit{uvm2} and detections in \textit{uvw1} are included in Table \ref{tab:phot} and plotted in Figure \ref{fig:lc}. 

\begin{table*}[ht]
\begin{center}
\caption{An example of the photometric observations of SN~2018ivc. A complete machine readable table is included in the online materials. }
\label{tab:phot}
\begin{tabular}{ccccccc}
\hline\hline
Observation Date & MJD & Phase & Source & Filter & Magnitude & Magnitude Error \\
UT &  & (day) &  &  & (mag) & (mag) \\
\hline
2018-11-13 02:16:15.16 & 58435.09 & -9.16 & CTIO-Prompt5 & Open & $\rm{<15.26}$ & - \\
2018-11-14 02:33:08.64 & 58436.11 & -8.14 & CTIO-Prompt5 & Open & $\rm{<15.29}$ & - \\
2018-11-15 02:24:00.86 & 58437.10 & -7.15 & CTIO-Prompt5 & Open & $\rm{<15.79}$ & - \\
2018-11-15 05:56:24.00 & 58437.25 & -7.00 & ZTF & g & $\rm{<19.82}$ & - \\
2018-11-15 07:29:16.80 & 58437.31 & -6.94 & ZTF & g & $\rm{<20.16}$ & - \\
2018-11-16 03:03:18.72 & 58438.13 & -6.12 & CTIO-Prompt5 & Open & $\rm{<15.66}$ & - \\
2018-11-17 02:50:24.57 & 58439.12 & -5.13 & CTIO-Prompt5 & Open & $\rm{<15.75}$ & - \\
2018-11-18 02:02:48.19 & 58440.09 & -4.16 & CTIO-Prompt5 & Open & $\rm{<15.57}$ & - \\
2018-11-19 02:39:18.43 & 58441.11 & -3.14 & CTIO-Prompt5 & Open & $\rm{<19.36}$ & - \\
2018-11-20 10:14:52.00 & 58442.43 & -1.82 & ATLAS & o & $\rm{<18.60}$ & - \\
2018-11-24 00:52:16.32 & 58446.04 & 1.79 & CTIO-Prompt5 & Open & 14.65 & 0.01 \\
2018-11-24 01:14:37.24 & 58446.05 & 1.80 & CTIO-Prompt5 & Open & 14.66 & 0.01 \\
2018-11-24 01:06:45.50 & 58446.05 & 1.80 & CTIO-Prompt5 & Open & 14.68 & 0.01 \\
2018-11-24 01:33:23.90 & 58446.06 & 1.81 & CTIO-Prompt5 & Open & 14.63 & 0.01 \\
2018-11-24 02:14:47.04 & 58446.09 & 1.84 & LCO LSC 1m & U & 14.35 & 0.02 \\
\hline
\end{tabular}
\begin{tablenotes}
The \textit{UVW1}, \textit{UVW2}, \textit{UVM2}, \textit{U}, \textit{B}, and \textit{V} filters are given in the Vega magnitude system; the \textit{g}, \textit{r}, and \textit{i} filters are given in the AB magnitude system.
\end{tablenotes}
\end{center}
\end{table*}

%--------- Spectroscopy ------------%
\subsection{Spectroscopy}\label{sec:obs_spec}
\begin{figure}[ht]
    \includegraphics[width=\columnwidth]{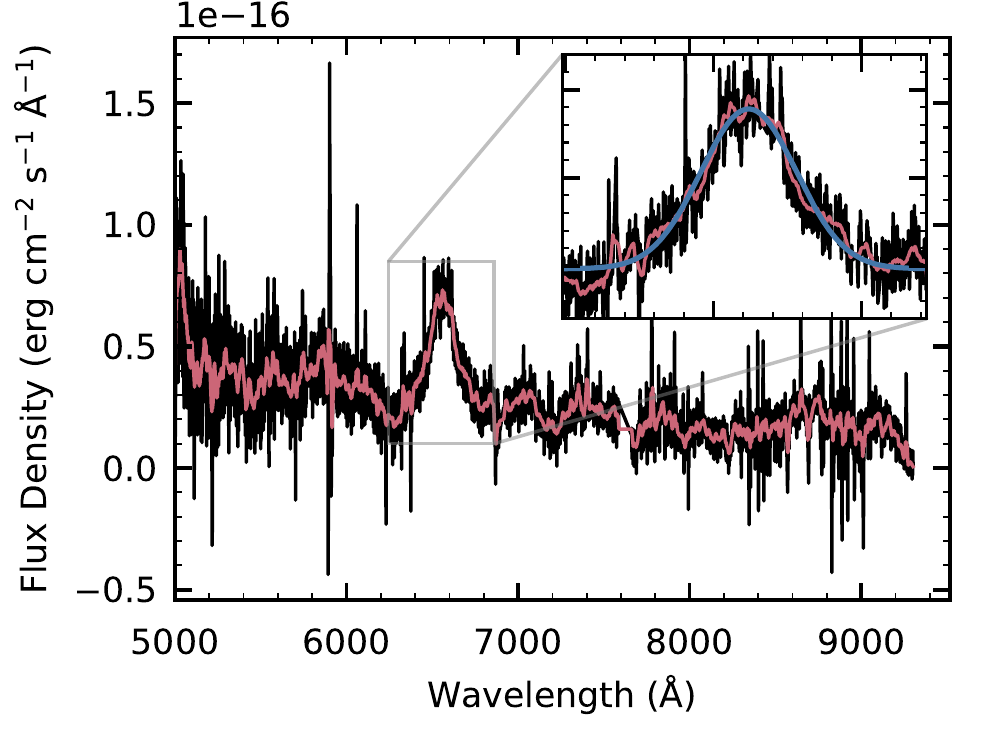}
    \caption{The Keck/DEIMOS nebular spectrum, observed on day 279 (black; pink is smoothed by 25 pixels) with a zoom-in on the \Ha{} region in the inset.
    A single Gaussian fit is shown in the inset in blue.
    Although the spectrum has a low signal-to-noise ratio, there is no evidence of multiple components in the \Ha{} feature.}
    \label{fig:nebular}
    \end{figure}

Spectroscopic observations from a variety of telescopes and instruments were obtained almost daily in the optical, starting within 5\,hr of discovery and continuing through day 35.
Less frequent monitoring with larger telescopes was performed until SN~2018ivc disappeared behind the Sun $\sim80$ days post-discovery. 
A single nebular spectrum was obtained with Keck/DEIMOS on day 279 (see Figure \ref{fig:nebular}).

\begin{figure*}[ht]
\begin{center}
\includegraphics[width=\textwidth]{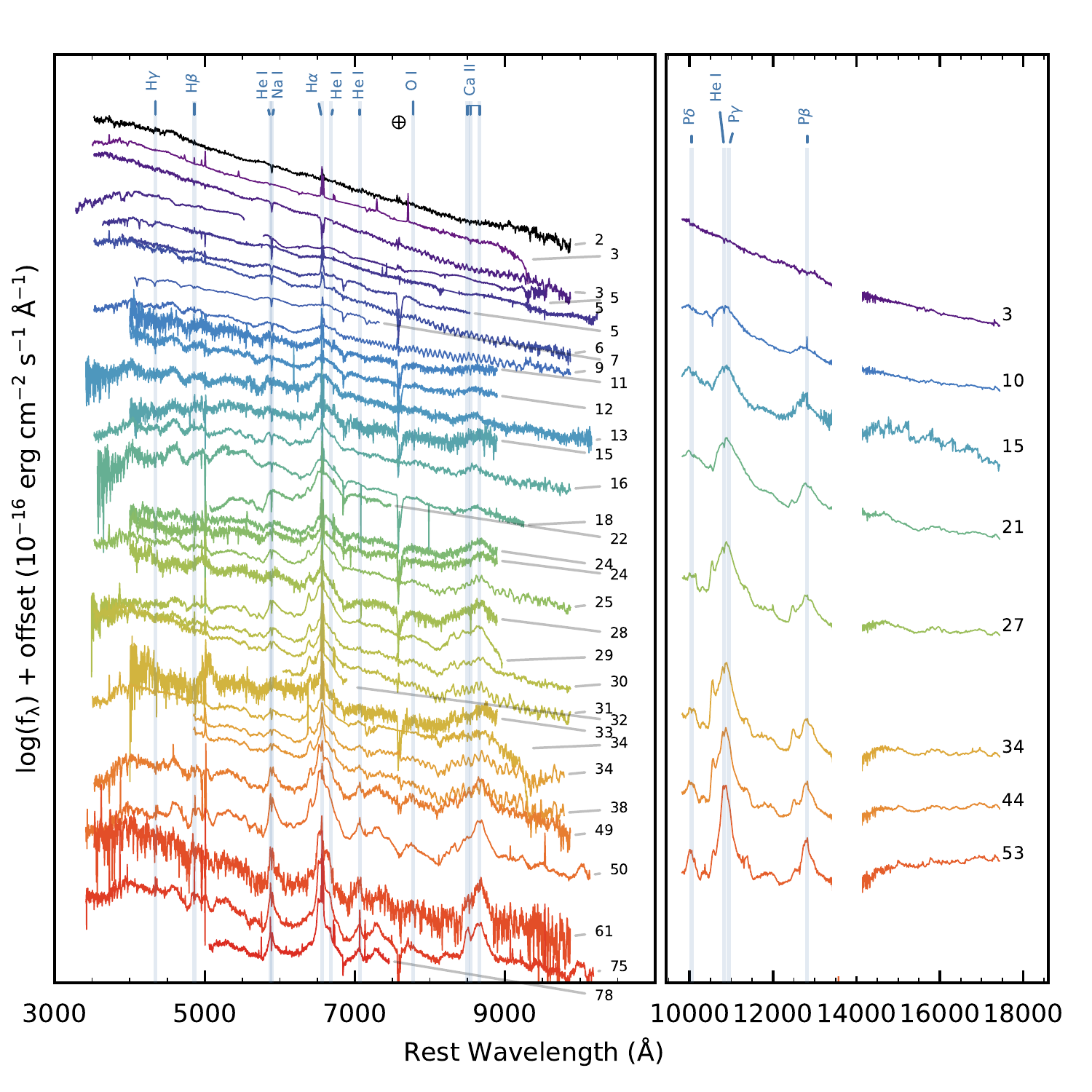}
\caption{\textit{Left}: the spectroscopic evolution of SN~2018ivc at optical wavelengths. 
SN~2018ivc shows strong emission lines and very little absorption. 
Imperfect background subtraction may create artificially blue continua and/or strong narrow lines in the broad \Ha{} emission.
The phase of each spectrum is given on the right.
High resolution observations are resampled with \texttt{SpectRes} \citep{2017Carnall} to 2\AA{} resolution to improve the signal-to-noise ratio.
\textit{Right}: the NIR spectroscopic evolution with the phase of the spectrum given on the right.
The strong telluric absorption between 13450 \AA{} and 14000 \AA{} is masked in all NIR spectra.
At the top of each panel prominent lines are identified, and in the left panel a prominent telluric feature is marked with a cross.
\label{fig:spec}}
\end{center}
\end{figure*}

Similarly, NIR observations began on day 3 and continued almost weekly through day 53. 
A selection of spectra is shown in Figure \ref{fig:spec} and all spectroscopic observations are listed in Table \ref{tab:spec}.
All of the spectra presented in this work were obtained at the parallactic angle to minimize atmospheric refraction \citep{1982filippenko}. 
These spectra will be made available on WISeREP\footnote{http://wiserep.weizmann.ac.il} \citep{Yaron12}.

Optical spectra were reduced using standard techniques, including bias subtraction, flat fielding, and cosmic-ray rejection.  
SN~2018ivc is embedded in NGC~1068; given this complex background, local sky subtraction was very important for the spectral extraction.  
Despite the care taken, some narrow emission lines are still visible in the final reduced spectra.
In these cases, after visual inspection of two-dimensional spectra and our highest resolution data (see Section~\ref{sec:SpecEvolve} for a detailed discussion), we believe these originate from the host galaxy.

Flux calibration was performed using standard-star observations.  
For the NIR spectra, the data were reduced in a similar manner as by \citet{Hsiao19}, using the standard ABBA technique; observations were taken of nearby A0V stars adjacent to the science exposures to facilitate telluric corrections and flux calibration \citep[e.g.,][]{Vacca03}.

\subsection{X-ray Observations}
The \cxo\ observed SN 2018ivc on 2018 Dec. 05.7 for 10.0\,ks (ObsID 20306) with the telescope aimpoint on the Advanced CCD Imaging Spectrometer (ACIS) S3 chip as part of a program to follow up possible X-ray detections from other facilities (PI: D.\ Pooley).  
The host galaxy, NGC 1068, has been observed many times previously with \chandra, often but not always with the High Energy Transmission Gratings (HETG) in place.  
For a straightforward comparison to our observation, we selected the longest ACIS-S3 observation with no grating in the \chandra\ data archive: ObsID 344 (PI: A.\ Wilson) began on 2000 Feb 21.7 and had an exposure time of 47.4\,ks.  
Data reduction was performed with the {\tt chandra\_repro} script, part of the {\tt Chandra Interactive Analysis of Observations (CIAO) software} \citep{2006SPIE.6270E..1VF}.  
We used {\tt CIAO} version 4.11 and calibration database (CALDB) version 4.8.3.  

\begin{figure*}[ht]
\includegraphics[width=\textwidth]{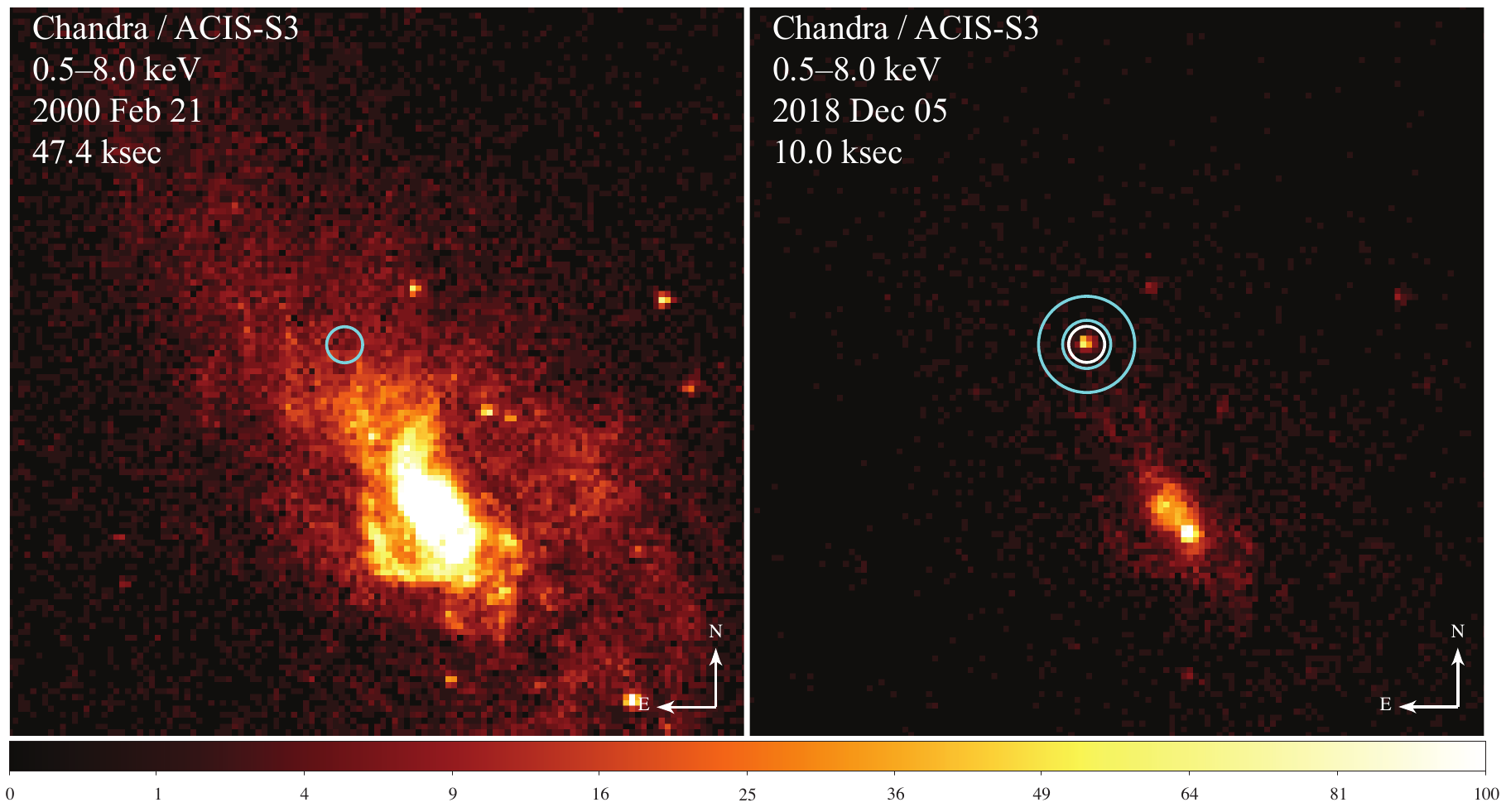}
\caption{Pre-SN ({\it left}) and post-SN ({\it right}) images of NGC~1068 taken with \chandra/ACIS-S3.  Each image is 1\arcmin\ on a side.  The white circle on the right is 1\farcs5 in radius centered on SN~2018ivc and is used to extract the source spectrum.  Note the non-uniform extended emission surrounding the SN. The cyan circle on the left (identical to the source region) and the cyan annulus on the right were each used to extract background spectra.  See text for details.}
\label{fig:xrayimage}
\end{figure*}

The source is clearly detected (Figure~\ref{fig:xrayimage}) with 207 total counts recorded in a 1\farcs5 radius source extraction region in the 0.5--8\,keV band.  
The background contribution to this X-ray flux is non-negligible but difficult to estimate given the non-spatially-uniform X-ray emission immediately surrounding the location of SN 2018ivc in its host galaxy. 
One estimate of the background comes from an annular background region of inner radius 2\arcsec\ and outer radius 4\arcsec\ centered on the SN.  
Based on the 80 counts in this background region, there are $192\pm14$ net counts from SN 2018ivc in the 0.5--8\,keV band.   

To assess the spectral properties of the SN and determine its X-ray flux, we extract and simultaneously fit source and background spectra in the 0.5--8\,keV band using {\tt Sherpa} \citep{2001SPIE.4477...76F} with the modified \citet{1979ApJ...228..939C} statistic cstat and the simplex optimization method. 
Our model components in all cases are absorbed hot plasmas (APEC model).  
We use the Tuebingen-Boulder Interstellar Medium absorption model \citep{2000ApJ...542..914W} with a minimum column density equal to the Galactic value of $\nH = \ee{1.55}{19}~\pcmsq$.   
We separately use two options for the background spectrum: the first is the annulus mentioned above in ObsID 20306 and the second is a 1\farcs5 radius circle (identical to the source extraction region) in ObsID 344.  

The first choice of background spectrum (annular extraction region from ObsID 20306) is well fit by two hot plasmas (temperatures of $kT_1 = 0.01$\,keV and $kT_2 = 0.85$\,keV) absorbed by a column of density \ee{3.2}{20}~\pcmsq.  
The second choice of background spectrum (circular extraction region from ObsID 344) is also well fit by two hot plasmas (temperatures of $kT_1 = 0.21$\,keV and $kT_2 = 0.95$\,keV) absorbed by a column of density \ee{1.6}{19}~\pcmsq.   
Using each of these background spectra separately, we fit the SN spectrum with an absorbed plasma.  

With the first choice of background, we obtain a good fit (reduced cstat of 0.63) with best-fit parameters for the SN of $\nH = \ee{(4.4\err{0.7}{0.9})}{22}~\pcmsq$ and $kT = 17\err{63}{7}$\,keV with the upper limit representing the model maximum.  
All uncertainties are 68\% confidence intervals.  
The intrinsic X-ray flux of the SN in this model is \ee{(8.2\pm0.8)}{-13}~\ergcms.  
For a distance of 10.1 Mpc, this corresponds to a luminosity of $\Lx = \ee{(1.0\pm0.1)}{40}~\ergsec$.

With the second choice of background, we also obtain a good fit (reduced cstat of 0.59) with best-fit parameters for the SN of $\nH = \ee{(3.6\err{0.8}{0.6})}{22}~\pcmsq$ and $kT = 43\err{37}{31}$~keV with the upper limit representing the model maximum.  
The intrinsic X-ray flux of the SN in this model is \ee{(7.5\pm0.7)}{-13}~\ergcms.  
For a distance of 10.1 Mpc, this corresponds to a luminosity of $\Lx = \ee{(9.2 \pm 0.9)}{39}~\ergsec$.

Although the different choices for background extraction give slightly different results, they are consistent with each other within the uncertainties.  
In each case, the reported fluxes are integrated from the unabsorbed models.  
Uncertainties on those fluxes are calculated as the 68\%-confidence bounds of the integrated, unabsorbed fluxes from Monte Carlo realizations (1000 samples) of the best-fit models, taking into account the uncertainties in the best-fit parameters (using the sample\_flux command in Sherpa) 

%%%%%%%%%%%%%%%%%%%%%%%%%%%%%%%%%%%%
%--------- Supernova Parameters ------------%
%%%%%%%%%%%%%%%%%%%%%%%%%%%%%%%%%%%%
\section{Properties of the Supernova and its Host Galaxy}\label{sec:snparams}
\subsection{Supernova Parameters}
The DLT40 survey identified SN~2018ivc on 2018 Nov. 24.07, 4.96 days after the last observation of the field on 2018 Nov. 19.11. 
After the DLT40 team's prompt reporting of the SN to TNS, the Asteroid Terrestrial-impact Last Alert System (ATLAS) identified a more recent nondetection in their data on 2018 Nov. 20.42 with a limiting magnitude of 18.6 mag in the orange-ATLAS filter.
As an explosion epoch we adopt 2018 Nov. $22.25 \pm 1.8$, the midpoint of the last nondetection (by ATLAS) and the first detection (by DLT40). 

We adopt Milky Way extinction values of $E(B-V)=0.0289\pm0.0004$ mag from \citet{Schlafly_2011}.
Unfortunately, SN~2018ivc exploded in a region of high extinction within NGC~1068, which prevented us from deriving the extinction in the host galaxy from the \NaI~D lines \citep{Poznanski2012}.
Instead, we estimate an extinction of $E(B-V)=0.5\pm0.15$ mag by matching the color evolution of SN~2018ivc over the first 20 days with that of other Type II SNe (SN~1980K: \citealt{1982barbon,1982buta,1983tsvetkov}; SN~1998S: \citealt{2000fassia,2000liu,2011li,2004pozzo}; SN~1996al: \citealt{2016benetti}; SN~2012A: \citealt{Tomasella2013}, SN~2013by: \citealt{Valenti2015}, SN~2013ej: \citealt{Valenti14}). 
While the light curves of these SNe have varying slopes, the colors are consistent during the first 20 days, which is why we choose this period to constrain the extinction. 
A conservative error of 0.15 mag is adopted to take into account the large uncertainty of the method. 
This is consistent with the extinction derived in the next section from spectroscopy of the parent \HII{} region and we adopt it as the host-galaxy extinction for our analysis.
Throughout this paper, unless otherwise noted, we use the extinction law of \citet{Cardelli1989} with $R_{V}=3.1$.

In Type IIP/IIL SNe, the amount of $\rm{{}^{56}Ni}$ synthesized in the explosion can be calculated from the luminosity of the SN after the fall from plateau ($\sim 80$--100 days post explosion), when the light curve is powered by the radioactive decay of $\rm{{}^{56}Ni\rightarrow {}^{56}Co\rightarrow {}^{56}Fe}$.
We measure the pseudo-bolometric luminosity of SN~2018ivc from the \textit{HST} observations on day 220.9 by first transforming the \textit{F555W} filter and the \textit{F814W} filter to the Landolt \textit{V} and \textit{I} filters, respectively, using the relations of \citet{2018harris}.
This first step is necessary as SN~1987A was not observed in the WFC3 \textit{F555W} and \textit{F814W} filters.
Following \citet{2008valenti}, we find the pseudo-bolometric luminosity by integrating the apparent magnitude at the effective wavelength of each filter, using Simpson's rule. 
We calculate the pseudo-bolometric luminosity for SN~1987A from the \textit{V} and \textit{I} filters in the same way.
Then, following \citet{2014spiro}, we calculate the nickel mass as $M({}^{56}{\rm Ni}) = 0.075\,M_{\odot} \times {L_{\rm 18ivc}(t)}/{L_{\rm 87A}(t)}$.
We find $M(\rm{{}^{56}Ni)=0.0056^{+0.0036}_{-0.0022}}\,M_{\odot}$.
Uncertainties in the nickel mass were calculated using a Monte Carlo simulation taking into account normal uncertainties in the explosion epoch, distance modulus, Galactic extinction, host galaxy extinction, and apparent magnitude of SN~2018ivc and SN~1987A.
We caution that to calculate this value we assumed that there was complete $\gamma$-ray trapping, that we could convert \textit{HST} filter magnitudes to Landolt filter magnitudes using relations derived from stellar spectral energy distributions (SEDs), and that SN~1987A and SN~2018ivc have the same SED.
The uncertainties associated with each of these assumptions are not included in the reported uncertainties.
%%%%%%%%%%%%%%%%%%%%%%%%%%%%%%%%%%%%
%--------- Host -------------%
%%%%%%%%%%%%%%%%%%%%%%%%%%%%%%%%%%%%
\subsection{Host Properties}\label{sec:host}
We searched in the ESO Science Portal\footnote{http://archive.eso.org/scienceportal/home} for MUSE integral field unit (IFU) observations of SN~2018ivc's host galaxy, NGC~1068. 
It was observed on 2014 Dec. 14 under program 094.B-0298(A), in four pointings which we combined for a total exposure time of 1180\,s.
All observations were reduced with the standard MUSE pipeline \citep{2014weilbacher} using default parameters through \texttt{reflex} \citep{2013freudling}.

For each spaxel, we performed a similar analysis to that of \cite{2014A&A...572A..38G,2016A&A...591A..48G, 2016galbany2}.
Briefly, using a modified version of {\sc STARLIGHT} (\citealt{2005MNRAS.358..363C,2016MNRAS.458..184L}, priv. comm.), we model the stellar component of the continuum by estimating the fractional contribution of simple stellar populations (SSP) from the \cite{2007ASPC..374..303B} base adding dust attenuation effects as a foreground screen with a \cite{1999PASP..111...63F} reddening law and $R_V = 3.1$. 
Our basis set is composed of 66 SSPs with 17 ages, ranging from 1\,Myr to 18\,Gyr, and four different metallicities (0.2, 0.4, 1.0, and 2.5 $Z_\odot$).

By subtracting the best SSP fit from each observed spectrum, we obtained a pure gas emission spectrum for each spaxel.
From the pure gas spectrum,  we estimated the flux of the most prominent emission lines after correcting for dust content with a correction derived from the Balmer decrement (assuming case B recombination; \citealt{2006agna.book.....O}, the same extinction law, and $R_V=3.1$). 
From the pure gas models for each spaxel, we create an extinction-corrected \Ha{} map.

\begin{figure*}[ht]
    \centering
    \includegraphics[width=\textwidth]{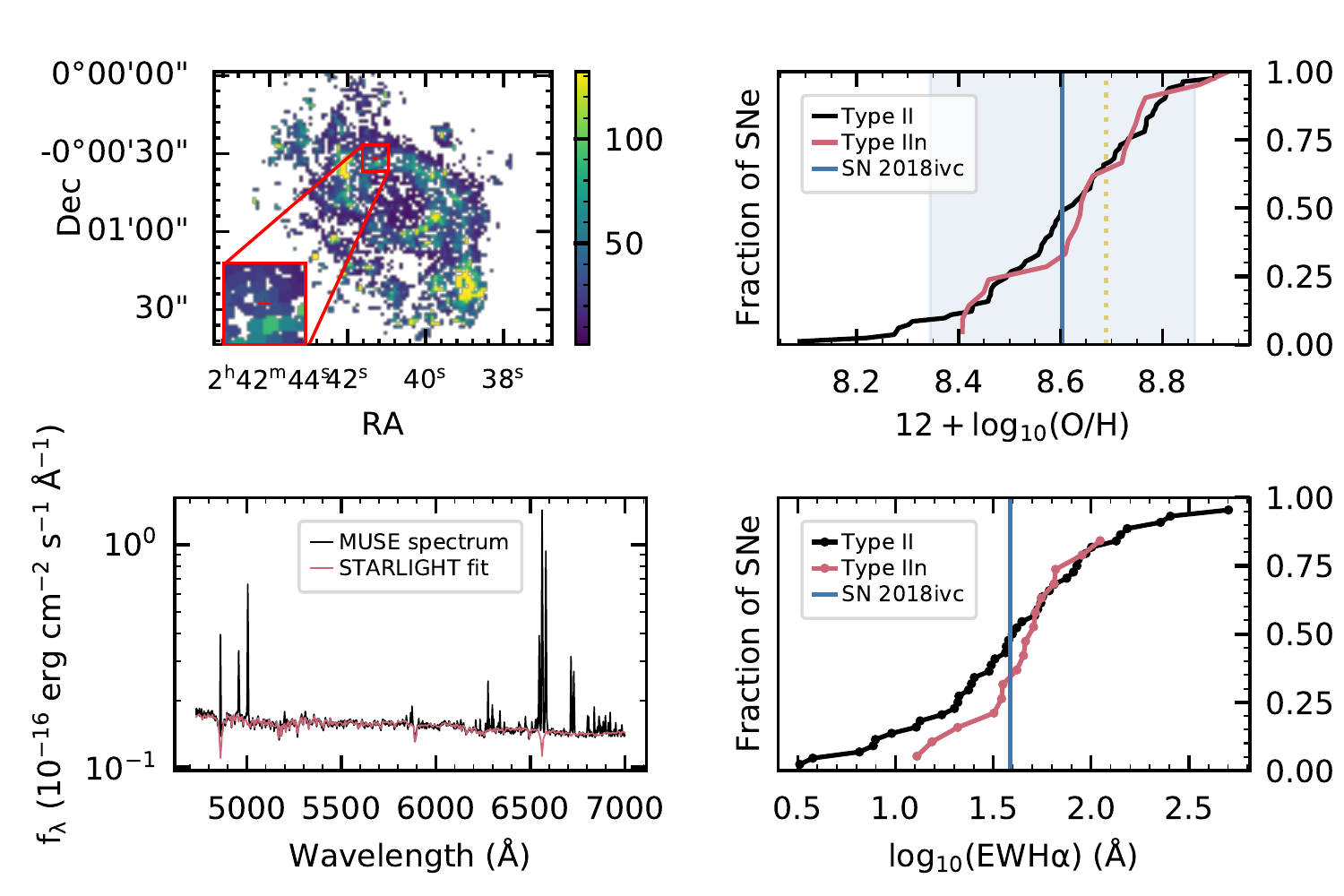}
    \caption{
    \textit{Upper left}: the map of the \HII{} regions in NGC 1068 identified by {\sc {\mbox HIIexplorer}}; the parent \HII{} region of SN~2018ivc is marked with a red point. 
    The map is colored by \Ha{} equivalent width (EW\Ha), which is a proxy for regions with young stellar populations.
    A close-up view of a 10 \arcsec region around the SN is shown in the inset.
    \textit{Lower left}: the MUSE spectrum of the parent \HII{} region of SN~2018ivc (black) and the STARLIGHT fit to the spectrum in pink.
    \textit{Upper right}: the cumulative fraction of SNe from the PISCO sample \citep{Galbany2018} as a function of metallicity for Type IIP/IIL SNe (black) and Type IIn SNe (pink).
    The metallicity of the parent \HII{} region of SN~2018ivc is marked with a blue vertical line and the uncertainty in this value is denoted with the light blue shaded region.
    Solar metallicity (8.69; \citealt{2009Asplund}) is plotted as a dotted yellow line.
    \textit{Lower right}: the cumulative fraction of SNe as a function of EW\Ha{} of their parent \HII{} region. The fraction of Type IIP/IIL SNe is shown in black, the fraction of Type IIn SNe in pink, and the EW\Ha{} of SN~2018ivc is marked with the blue vertical line. }
    \label{fig:host}
\end{figure*}

We use our extinction-corrected \Ha{} map to identify the \HII{} region containing SN~2018ivc (its ``parent'' region) and derive environmental and progenitor properties.
Following \cite{2016MNRAS.455.4087G,2018ApJ...855..107G}, the \Ha{} maps are used to select star-forming \HII{} regions across the galaxy with a modified version of {\sc {\mbox HIIexplorer}}\footnote{http://www.caha.es/sanchez/HII\_explorer/} \citep{2012A&A...546A...2S}, a package that detects clumps of higher intensity in a map by aggregating adjacent pixels. 
This procedure selected 1801 \HII{} clumps with an average radius of 140\,pc. 
Once the \HII{} regions were identified, the same analysis described above was performed on the extracted spectra. 
The observed spectrum and the best-fit STARLIGHT spectrum of the \HII{} region containing SN~2018ivc can be seen in the bottom-left panel of Figure \ref{fig:host}.
From this analysis and using the \citet{Cardelli1989} extinction law, we find $E(B-V)=0.37\pm0.04$ mag for the parent region of SN~2018ivc, consistent with the value derived in the previous section.

We use oxygen abundance as a proxy for metallicity as oxygen is produced at the beginning of the enrichment process by massive stars and, by mass, comprises 50\% of the heavy elements in the Universe \citep{2012Lopez}. 
Using the pure gas emission spectrum of the \HII{} region, we determine metallicity from the O3N2 empirical calibrator  \citep{2004MNRAS.348L..59P}.
This same spectrum is used to find the equivalent width of \Ha{} (EW\Ha{}).
The upper-left panel of Figure \ref{fig:host} shows the EW\Ha{} values for each \HII{} region in NGC~1068. 
We find the metallicity to be $\rm{12+log_{10}(O/H) = 8.6\pm0.26}$ and the EW\Ha{}$=38.58\pm 0.26$ \AA.  
These host properties will be put in the context of other SNe in Section \ref{sec:Discussion}.

%--------- Evolution ------------%
\section{Supernova Evolution}\label{sec:evolve}
The detailed evolution of SN~2018ivc was caught in the almost daily photometric and spectroscopic observations. 
These revealed a rapidly evolving SN with a steeply declining light curve and spectra with broad emission.

%--------- Photometric Evolution ------------%
\subsection{Light-Curve Evolution}
\begin{figure}[ht]
    \centering
    \includegraphics[width=\columnwidth]{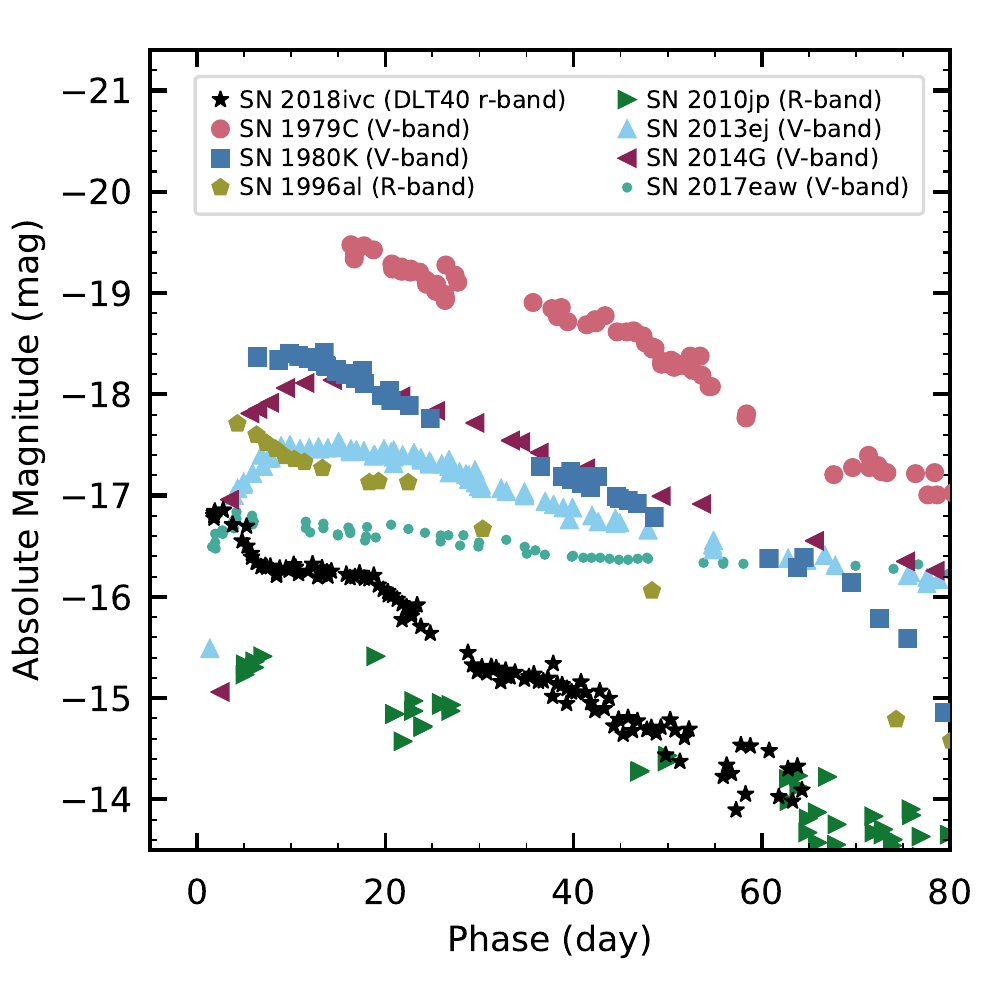}
    \caption{The light curve of SN~2018ivc (black stars; DLT40 \textit{r} band) compared to a sample of well-studied SNe.
    SNe with a variety of slopes are represented with an emphasis on Type IIL-like SNe.
    The best-observed filter is shown for each SN.
    SN~1979C \citep[pink circles, \textit{V} band; ][]{1980balinskaia, 1981devaucouleurs, 1982barbon} and SN~1980K \citep[blue squares, \textit{V} band;][]{1982barbon,1982buta,1983tsvetkov} represent the historical Type IIL-like class. 
    SN~2017eaw \citep[sea foam green points, \textit{V} band;][]{Szalai19} represents the Type IIP-like SNe. 
    SN~2013ej \citep[light blue triangles, \textit{V} band;][]{Valenti14} and SN~2014G \citep[maroon arrows, \textit{V} band;][]{2016terreran} are transitional Type IIL-like objects, with a clear fall from the plateau onto the radioactive decay tail.
    We plot two SNe that will be discussed later in the text in comparison to SN~2018ivc: SN~2010jp \citep[green arrows, \textit{R} band;][]{2012smith} and SN~1996al \citep[yellow pentagons, \textit{R} band;][]{2016benetti}. }
    \label{fig:lc_comp_other}
\end{figure}

The light curve of SN~2018ivc rose rapidly, peaking around day 3, only 5 days after the last nondetection.
It then showed a rapid, linear decline typical of Type IIL-like SNe.
Figure \ref{fig:lc_comp_other} displays a comparison of SN~2018ivc and other well-studied SNe with a variety of slopes.
While the global trend is linear with a similar decline rate to SNe 1979C and 1980K, our high-cadence observations quickly revealed that there are, in the first 50 days, several changes in the slope rarely seen in other Type IIL-like SNe.
SN~2018ivc is also relatively faint compared to other SNe with similar slopes. 

The light curve of SN~2018ivc changes slope approximately every 10 days over the first 30 days. 
We fit a continuous piecewise linear function to the \textit{r}-band observations, leaving the slopes, initial intercept, and breakpoints as free parameters (8 free parameters).  
This fit is shown in Figure \ref{fig:lc}.
The initial decline ending at day 
$\rm{ 7.57 \pm 0.41}$ has a slope of $\rm{ 0.1068 \pm 0.0094}$ mag $\rm{day^{-1}}$.
This is followed by a plateau until day
$\rm{18.07 \pm 0.45}$ with a slope of $\rm{0.0056 \pm 0.0031}$ mag $\rm{day^{-1}}$. 
The light curve begins to decline steeply on day 
$\rm{ 18.07}$, with a slope of $\rm{ 0.0811\pm 0.0077}$ mag $\rm{day^{-1}}$.
After day 
$\rm{27.55 \pm1.37 }$, the slope changes again to $\rm{0.0355 \pm 0.0014}$ mag $\rm{day^{-1}}$. 
At no point is the slope of SN~2018ivc consistent with the slope expected for cobalt decay (0.008 mag $\rm{day^{-1}}$ in $R$). 
Searching among other Type IIP/IIL SNe, we were only able to find one light curve in the literature that resembled the evolution (although $\sim 1$ mag brighter): that of SN~1996al (see Section \ref{sec:96al}).

We searched the DLT40 difference images in the months leading up to SN~2018ivc for a pre-explosion outburst similar to SN~2009ip ($M_r \approx -14.5$ mag; \citealt{2011foley, Mauerhan13, 2013pastorello, 2014marguitti}).
However, these observations do not reveal any hint of a pre-outburst eruption, with $\sim90$ images of the field take in the $\sim150$ days prior to the SN explosion, with a typical limiting magnitude of $r \approx 19.3$  ($M_r \approx -12$ mag). 

%%%%%%%%%%%%%%%%%%%%%%%%%%%%%%%%%%%%
%---Spectroscopic Evolution -------%
%%%%%%%%%%%%%%%%%%%%%%%%%%%%%%%%%%%%
\subsection{Spectroscopic Evolution}\label{sec:SpecEvolve}
The spectroscopic evolution of SN~2018ivc shows strong H and \HeI{} emission and very shallow (if any) absorption. 
The full evolution can be seen in Figure \ref{fig:spec}, where prominent spectroscopic features are labeled.

The optical and NIR spectra obtained at 2.04 and 2.95 days (respectively) show a featureless blue continuum.
By day 5, broad \Ha{} and \HeI{} $\lambda5876$ emission begin to develop with similar profiles.
While the presence of \Hb{} in absorption cannot be completely ruled out (because of possible contamination from other lines), an absorption component is clearly missing from both \Ha{} and \HeI{} $\lambda5876$.
This is typical of Type IIn SNe and some Type IIL-like SNe \citep{2014gutierrez}.
Although the physical mechanism for the lack of absorption is still unclear, possible explanations include a low-density (possibly low-mass) envelope and scattering off of CSM \citep{1996schlegel}. 

The Ca~II $\lambda\lambda8498$, 8542, 8662 emission triplet begins to develop at day 12 and strengthens through day 78.
On day 30 the \HeI{} $\lambda7065$ line begins to show in emission.
By day 75, the \HeI{} $\lambda6678$ line is visible in the red wing of the \Ha{} emission; this feature may be blended in earlier spectra.
The shape of the emission around \Ha{} in the two spectra from days 75 and 78 is very boxy, which we will discuss more in later sections.
We obtained a low signal-to-noise ratio (S/N) spectrum at day 279 which shows a broad \Ha{} profile, similar to that seen on day 78.

The NIR spectra are dominated by hydrogen and \HeI{} emission features.
The blue continuum at day 3 gives way to Paschen emission features by day 10.
The \HeI{} $\lambda10830$ line is blended in the \Pg{} feature.
In both the optical and the NIR, the widths of the emission lines decrease with time as the speed of the ejecta decreases. 

%%%%%%%%%%%%%%%%%%%%%%%%%%%%%%%%%%%%
%--------- Narrow lines-------------%
%%%%%%%%%%%%%%%%%%%%%%%%%%%%%%%%%%%%
The spectra of SN~2018ivc show potential narrow emission lines on top of the broad \Ha{} emission.
This region is heavily contaminated by the host-galaxy emission which is challenging to separate from the \Ha{} SN emission in all but the highest resolution observations.
We searched for a SN component in our highest resolution spectrum: the MMT/Binospec observation from 2018 Dec. 14.
In this spectrum we fit a Gaussian profile to night-sky lines, galaxy emission lines, and \Ha{}.
We find the full width at half-maximum intensity (FWHM) of the narrow \Ha{} emission ($\sim130$ km $\rm{s^{-1}}$) to be consistent with that of the other galaxy emission lines ([N~II] $\lambda\lambda 6548$, 6583; [S~II] $\lambda\lambda 6716$, 6731).

\begin{figure*}[ht]
    \centering
    \includegraphics[width=\textwidth]{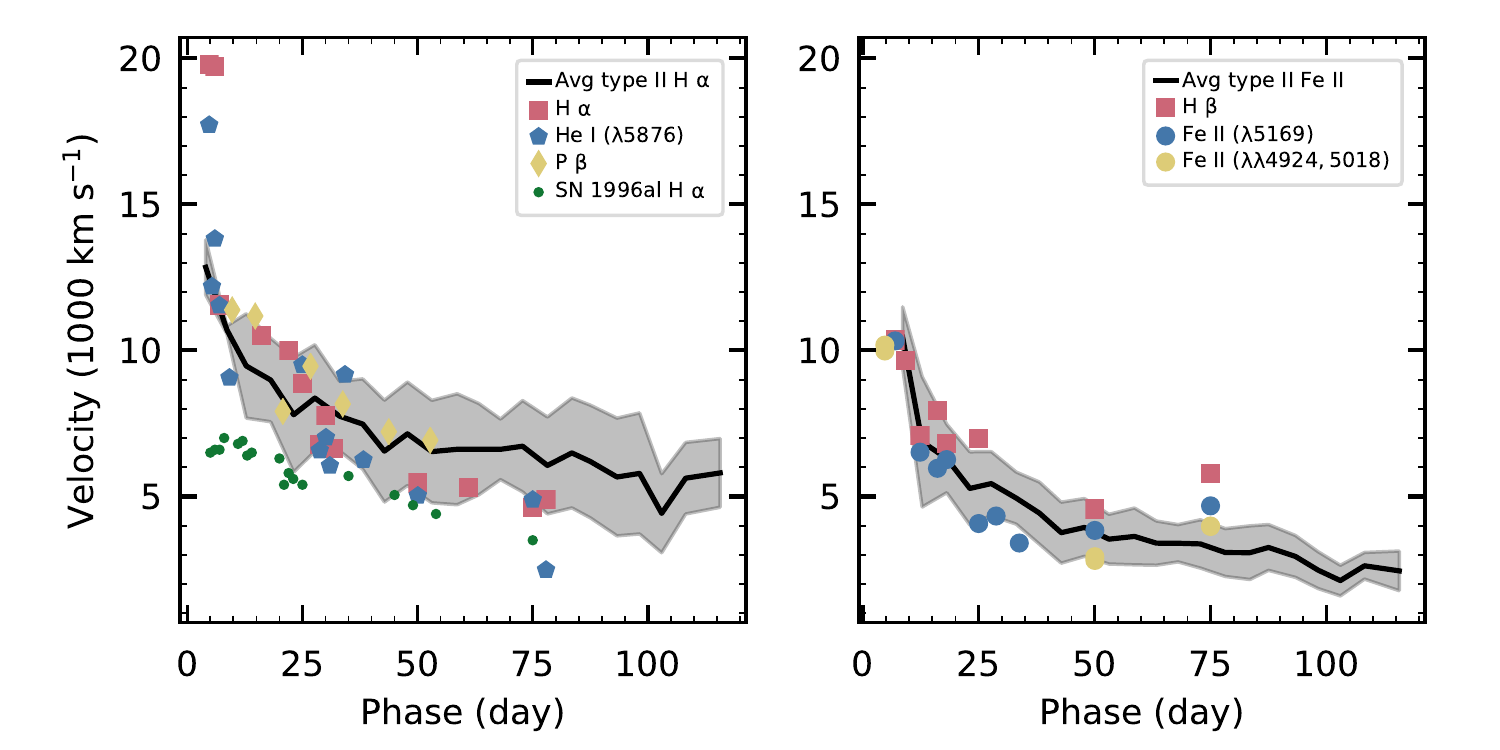}
    \caption{\textit{Left}: the expansion velocity evolution of SN~2018ivc found using the HWHM of emission lines \Ha{} (pink squares), \HeI{} (blue pentagons), and \Pb{} (yellow diamonds).
    The \Ha{} velocity of SN~1996al is shown as green circles. 
    The velocity of SN~2018ivc is similar to that of other Type IIP/IIL SNe,
    although it evolves more rapidly than most, including SN~1996al.
    \textit{Right}: the expansion velocity evolution of SN~2018ivc found using the minimum wavelength of the absorption features \Hb{} (pink squares) and \FeII{} (blue and yellow circles).
    The average \Ha{} velocity (left) and \FeII{} (right) of the \citet{Gutierrez2017} sample of Type IIP/IIL SNe is shown in black with the standard deviation in gray.  
    \label{fig:velocity}}
\end{figure*}

%%%%%%%%%%%%%%%%%%%%%%%%%%%%%%%%%%%%
%--------- velocities -------------%
%%%%%%%%%%%%%%%%%%%%%%%%%%%%%%%%%%%%
We measure the velocity evolution using the half width at half-maximum intensity (HWHM) of Gaussian fits to the \Ha{}, \HeI{} $\lambda5876$, and \Pb{} emission lines and the minimum of the \FeII{} and \Hb{} absorption features. 
The velocities are shown in Figure \ref{fig:velocity}, with velocities measured from emission features in the left panel and velocities measured from absorption features in the right panel.
We note that while there is asymmetry, substructure, and host contamination in individual features, the trends are consistent across features, giving us confidence in the overall velocity evolution.
However, we caution that there may be significant scatter in individual measurements.
The \Ha{} velocity can be compared to the average of 112 Type IIP/IIL SNe, calculated by \citet{Gutierrez2017}.
We find that the velocities are similar to those found in Type IIP/IIL SNe and higher than the velocities of SN~1996al.
However, the evolution is more rapid, with the initial velocity higher than average and the final velocities lower than average.
The \FeII{} lines are also broadly consistent with the average Type IIP/IIL evolution although there is considerable scatter. 
We simultaneously fit \FeII{} $\lambda\lambda 4924$, 5018, but we were unable to simultaneously fit \FeII{} $\lambda\lambda 4924$, 5018, 5169 and therefore fit \FeII{} $\lambda5169$ separately. 
Velocities from both fits are shown in Figure \ref{fig:velocity}.

\begin{figure*}[ht]
\begin{center}
\includegraphics[width=\textwidth]{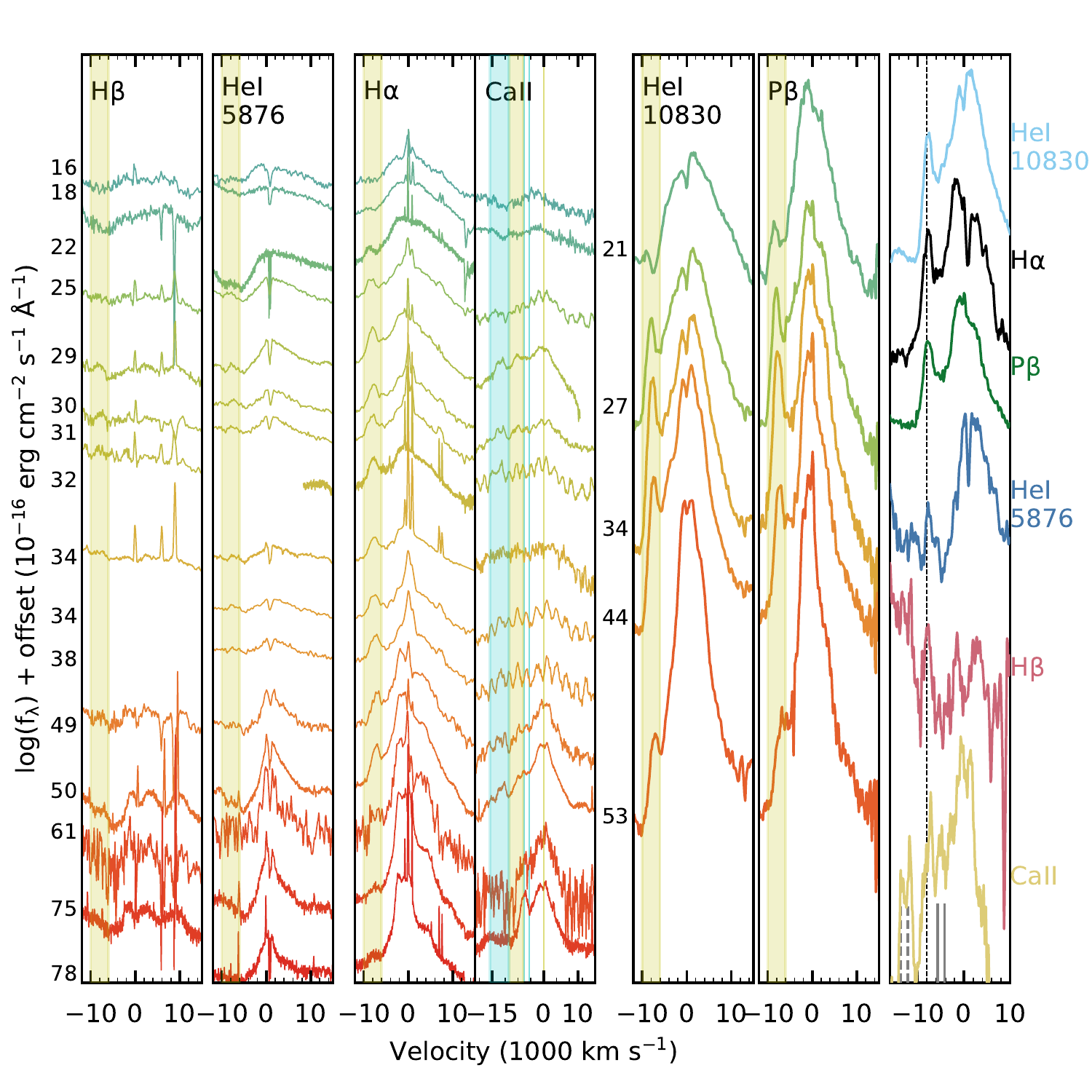}
\caption{The evolution of the high-velocity (HV) features in hydrogen (\Ha{}, \Hb{}, \Pb{}), helium ($\lambda 5876$, $\lambda 10830$), and the \CaII{} triplet from day 18 through day 78 in the optical and day 21 through day 53 in the NIR.
The feature is most prominent in \Ha{}, \Pb{}, and \HeI{} $\lambda 10830$.
The panels are labeled in the upper-left corner with the feature they present. 
The yellow shaded region marks $-10,000$ to $-6000$ \kms. 
The phases of the optical spectra are marked to the left of the first panel, and the NIR phases are marked to the left of the \HeI{} $\lambda 10830$ panel.
The \CaII{} panel is centered on the $\lambda8662$ component of the triplet (marked in yellow at both rest wavelengths and shaded in yellow at HV). 
The two blue components are marked at rest wavelength with vertical cyan lines and at high velocity with a shaded cyan region. 
The last panel shows the optical features on day 30 and the NIR features on day 34 for all emission lines in which the HV feature is seen. 
The HV feature is marked with a dashed line at $-8000$ \kms{} and each emission feature is labeled to the right of the panel. 
The bluest two \CaII{} triplet features that are offset in velocity space are marked with a solid line gray at the velocity corresponding to their rest wavelengths and dashed gray line for the HV features.
\label{fig:HV}}
\end{center}
\end{figure*}

%%%%%%%%%%%%%%%%%%%%%%%%%%%%%%%%%%%%
%--------- blue bump -------------%
%%%%%%%%%%%%%%%%%%%%%%%%%%%%%%%%%%%%
On day 18 an emission feature begins to develop in the blue wing of the emission-line profile of \Ha{}, \Hb{}, \HeI{} $\lambda 5876$, \CaII{} triplet, \HeI{} $\lambda 10830$, and \Pb{} (see the first five panels of Figure \ref{fig:HV}).
This feature gains strength until day $\approx 35$, after which it fades until it is barely visible on day 78.
We identify the emission line that appears blueward of the rest wavelength of these features as a high-velocity (HV) component owing to its presence at the same velocity in each feature (see right panel of Figure \ref{fig:HV}). 
From the blueshifted peak of the emission we find the velocity of the emitting material to be $\sim9000$ \kms{} at day 18, slowing down to $\sim7000$ \kms{} by day 78.

%%%%%%%%%%%%%%%%%%%%%%%%%%%%%%%%%%%%
%--- Analysis  -------%
%%%%%%%%%%%%%%%%%%%%%%%%%%%%%%%%%%%%
\section{Analysis}\label{sec:analyze}
%%%%%%%%%%%%%%%%%%%%%%%%%%%%%%%%%%%%
%---Compare to 1996al -------%
%%%%%%%%%%%%%%%%%%%%%%%%%%%%%%%%%%%%
\subsection{Comparison to 1996al}\label{sec:96al}
\begin{figure*}[ht]
\begin{center}
\includegraphics[width=\textwidth]{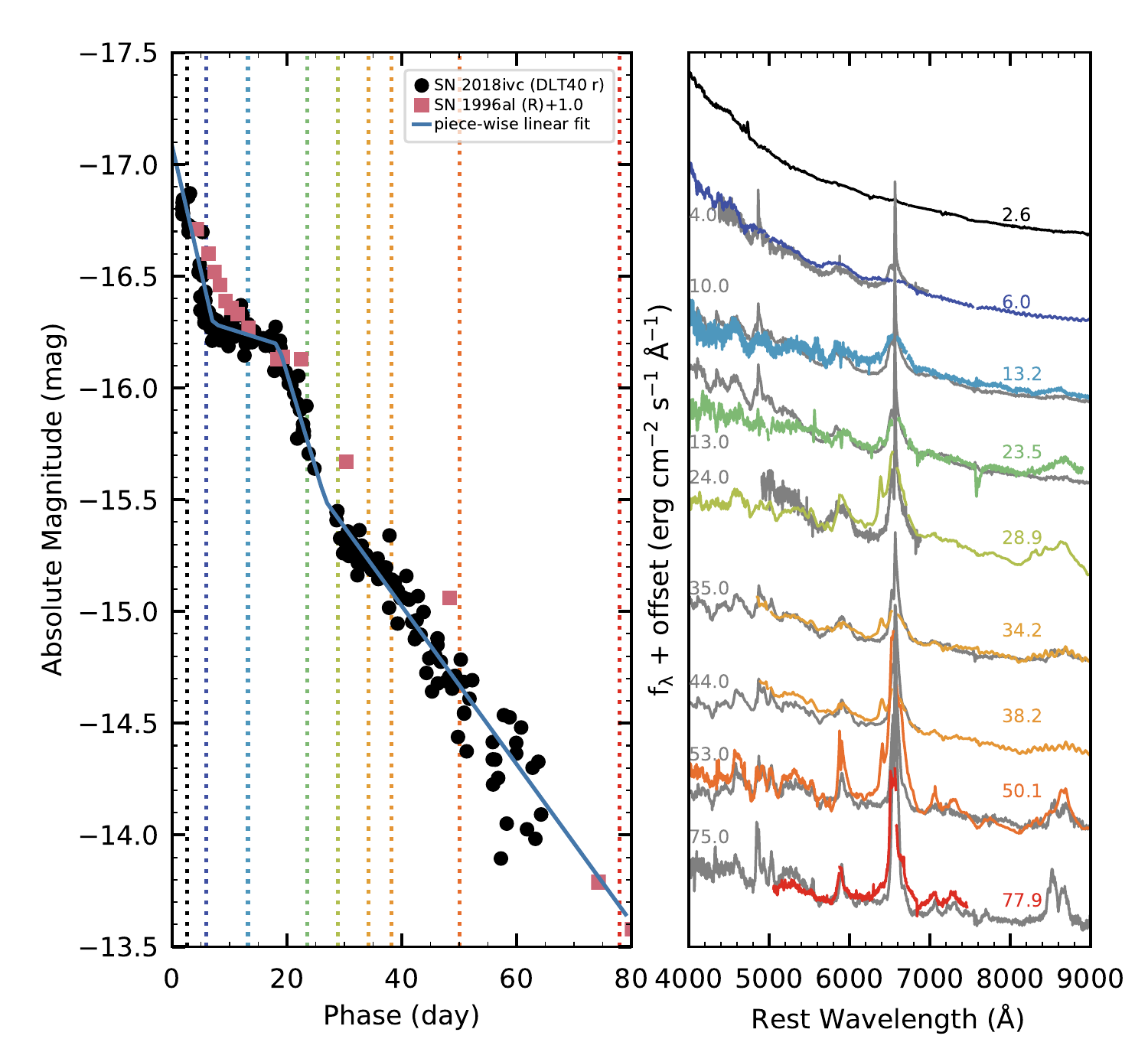}
\caption{Comparison of the photometric and spectroscopic evolution of SN~2018ivc to the similar event SN~1996al.
\textit{Left:} The \textit{r}-band light curve of SN~2018ivc (black) compared with the \textit{R}-band light curve of SN~1996al (pink). 
The light curve of SN~1996al has been shifted down by 1.0 mag to align with the SN~2018ivc light curve and the explosion epoch is set to 1996 July 19.
The piecewise linear fit of the SN~2018ivc light curve is shown in blue.
The dashed vertical lines correspond to the phase of the spectra plotted on the right.
\textit{Right}: The spectroscopic evolution of SN~2018ivc (colored spectra) compared with that of SN~1996al (gray spectra). 
The phases of SN~2018ivc are given on the right side of the figure in color and the phases of SN~1996al are given on the left side in gray. 
The spectra of SN~1996al have been offset to align with those of SN~2018ivc.
Although SN~1996al does not show the high-velocity features and SN~2018ivc does not have narrow lines (the host contamination visible in Figure \ref{fig:spec} has been masked out of the spectra of SN~2018ivc), SN~2018ivc and SN~1996al are spectroscopically similar with strong \HeI{} and hydrogen features in emission.
\label{fig:lc_comp}}
\end{center}
\end{figure*}

Figure \ref{fig:lc_comp} (left panel) shows the DLT40 \textit{r}-band light curve of SN~2018ivc in black compared with the \textit{R}-band light curve of SN~1996al in pink, shifted by 1.0 mag. 
Since there are no stringent detection limits to constrain the explosion epoch for SN~1996al, we use an explosion epoch of 1996 July 19, which is 18 days later than the reference epoch suggested by \citet{2016benetti}. 
They compare the light curve to that of other Type IIL-like SNe and estimate 1996 July 1 as the \textit{V}-band maximum, which they adopt as the reference epoch. 
However, given the similarity between SN~1996al and SN~2018ivc, using the latter to constrain the explosion of SN~1996al may be more appropriate.
While SN~2018ivc is very well sampled, the sparse sampling of SN~1996al, especially after the initial decline, makes a detailed comparison challenging.
Nevertheless, the light curve of SN~1996al does show a relatively steep decline, at a similar rate as SN~2018ivc that, like SN~2018ivc, is interrupted by a short plateau.

A spectroscopic comparison of SN~1996al (gray) and SN~2018ivc (color) is shown in the right panel of Figure \ref{fig:lc_comp}.
As described above, the phases are with respect to a reference time of 1996 July 19 for SN~1996al.
There is broad agreement in the spectral evolution of both SNe, with H and \HeI{} emission lines dominating the spectra and little to no absorption.
The width of the emission features of SN~1996al are smaller than those of SN~2018ivc, implying a higher ejecta velocity for SN~2018ivc.
This is quantified in the left panel of Figure \ref{fig:velocity}, which shows that the velocity of \Ha{} is slower in SN~1996al than in SN~2018ivc and that the \Ha{} velocity declines more rapidly than that of SN~1996al.

Given the overall similarity between SN~2018ivc and SN~1996al, we will here summarize the main results of \cite{2016benetti} for SN~1996al and compare these two objects more closely.
SN~1996al was identified as a transition SN between a Type IIL-like and Type IIn with a linearly declining light curve and a week-long plateau starting around day 15.
The spectra show broad hydrogen and \HeI{} emission with narrow P Cygni lines superimposed. 
From ground-based pre-explosion \Ha{} images, a 25 \msun{} luminous blue variable (LBV) progenitor was identified.
Light-curve modeling of the 15-year evolution showed that the light curve is dominated by ejecta-CSM interaction.
The linear decline of the light curve, the velocity evolution, and the evolution of the \Ha{} flux a year after explosion indicated a low ejecta mass and that the ejecta were predominantly helium.
\citet{2016benetti} interpret the similarity in shapes of the hydrogen and helium emission-line profiles as evidence that both these lines originate in the interaction region during at least the first 50 days of evolution. 
The low velocities and multicomponent emission lines point to a dense, asymmetric CSM, while the narrow P Cygni lines indicate a patchy, symmetric CSM at a larger radius.
The asymmetric CSM is confirmed in a multicomponent \Ha{} profile in a nebular spectrum from day 142.

Given the peculiarity of the light curve and spectroscopic evolution of SN~2018ivc, the qualitative similarity to SN~1996al during the first 80 days of evolution (before it disappeared behind the Sun) is striking.
The light curves decline at similar rates and show similar changes in slope, including a short plateau.
Additionally, their spectra are dominated by emission lines from the same species.
The steeply declining light curve and short plateau of SN~2018ivc could be a sign of a low mass of hydrogen in its ejecta.
\citet{2016benetti} model the light curve of SN~1996al and find that the light from the ejecta is significantly fainter than the light from the CSM.
The fact that SN~2018ivc is less luminous than SN~1996al indicates that the majority of the light may be coming from the SN ejecta, although the frequent changes in slope demonstrate that CSM-ejecta interaction is significant at some phases.

Additionally, the similarity of the hydrogen and \HeI{} emission profiles indicates that they originate in the same part of the ejecta and that perhaps, like SN~1996al, the ejecta of SN~2018ivc are predominantly helium.
While multipeaked emission profiles are not observed in the early-time spectra of SN~2018ivc, asymmetry is seen spectroscopically in the HV feature that appears around day 18.
The broad \Ha{} profile in the day 279 spectrum of SN~2018ivc does not show multiple components.
It is possible the SN~2018ivc has a multi-component \Ha{} profile, like the one clearly visible in SN~1996al, but that the two components are blended to the point of being indistinguishable from a single profile.
Given the similarity between SN~2018ivc and SN~1996al, it is possible that they have similar progenitors as we will discuss in Section \ref{sec:Discussion}.

%%%%%%%%%%%%%%%%%%%%%%%%%%%%%%%%%%%%
%---HST Progenitor Search -------%
%%%%%%%%%%%%%%%%%%%%%%%%%%%%%%%%%%%%
\subsection{Progenitor from {\sl HST\/} Pre-Imaging}
High-resolution images taken prior to explosion can be used to identify and characterize the properties of a SN progenitor \citep[e.g.][]{2011crockett, 2010eliasrosa, 2011eliasrosa, 2010fraser, 2011fraser,  2014fraser, 2012kochanek, 2007li, 2008mattila, 2005maund, 2009maund, 2013maund, 2014maund2, 2004smartt, 2009smartt, 2013tomasella, 2014maund1, 2017kochanek, 2018kilpatrick, 2019vandyk,VanDyk2017}.
We located pre-explosion {\sl HST} observations in the Mikulski Archive for Space Telescopes (MAST) and analyzed them for the presence of a progenitor. 
The SN site is located in Advanced Camera for Surveys/Wide Field Channel (ACS/WFC) data in bands F658N and F814W from program GO-9788 (PI: L.~Ho) and in F550M from GO-9503 (PI: N.~Nagar), as well as Wide Field Planetary Camera (WFPC2) images in F606W from both GO-5479 (PI: M.~Malkan) and GO-8597 (PI: M.~Regan) and in F450W from GO-11128 (PI: D.~Fisher); see Table~\ref{tab:HST}. 

\begin{figure*}[ht]
    \centering
    \includegraphics[width=\textwidth]{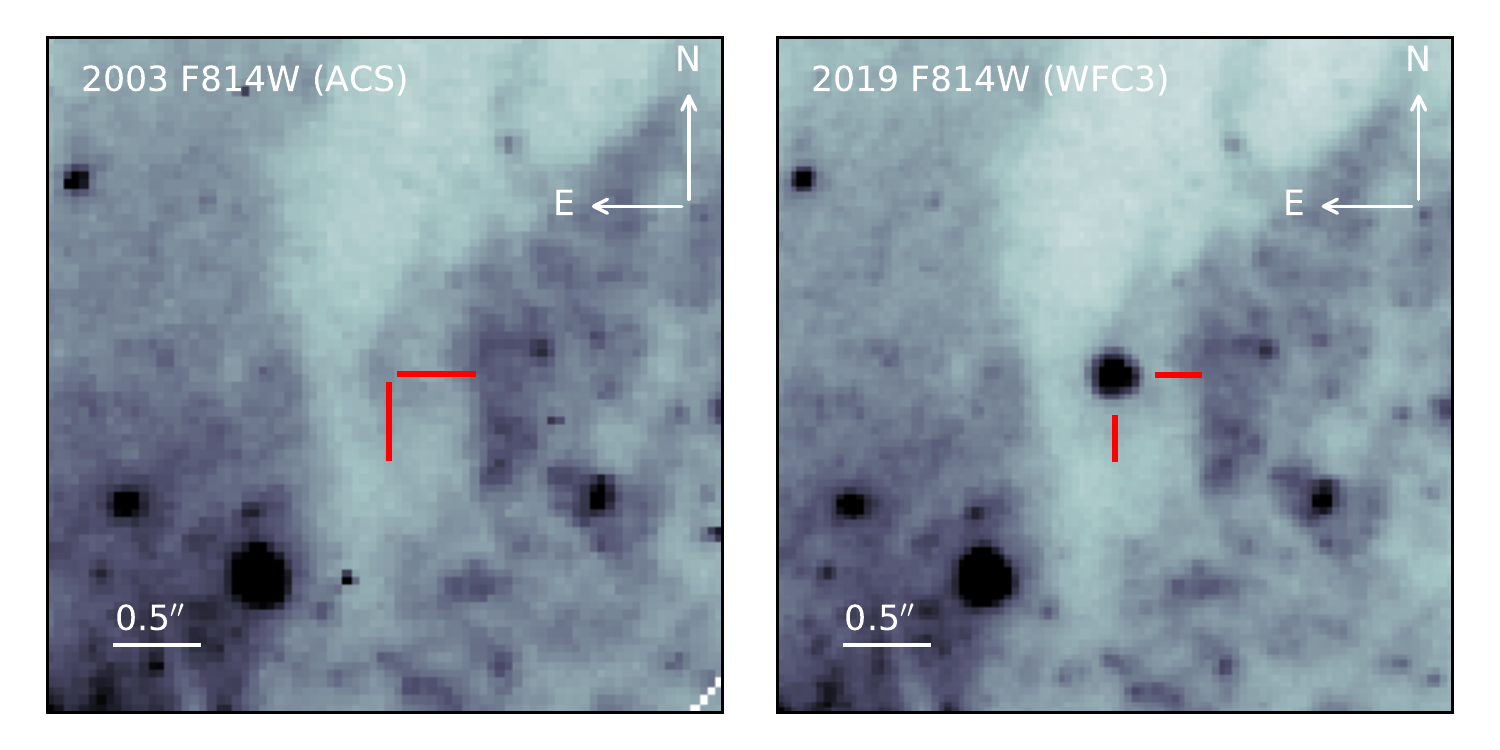}
    \caption{\textit{Left:} Pre-explosion {\sl HST\/} ACS/WFC image from 2003 Oct. 26 at F814W of NGC~1068 containing the site of SN~2018ivc. 
    No source is identified at the SN location, indicated by the red ticks.
    \textit{Right:} {\sl HST\/} WFC3/UVIS image at F814W of SN 2018ivc, obtained on 2019 July 1.
    The SN is indicated with the red ticks.}
    \label{fig:snsite}
\end{figure*}

To precisely pinpoint the SN location in the archival data, we subsequently obtained higher spatial resolution {\sl HST\/} images of the SN with the Wide Field Camera 3 (WFC3) on 2019 July 1. 
We identify 69 sources in common between between the F814W WFC3 image of the SN to the GO-9788 F814W ACS exposure and use these identify the SN location in the pre-explosion ACS image.
To do this, we randomly select 34 sources and use these to compute the astrometric transformation from the WFC3 image to the ACS image the \texttt{PyRAF} task \texttt{geomap}. 
We then measure the location of the SN in the pre-explosion image using the \texttt{PyRAF} task \texttt{geoxytran} and the astrometric transformation from \texttt{geomap}.
To understand the error introduced by the sources used to find the astrometric transformation, we calculate the root-mean-square (RMS) uncertainty of the 35 stars not used to determine the transformation.
We repeat this process 1000 times and find the SN to be located at pixel ($\rm{3546.990 \pm 0.011 \pm 0.062}$, $\rm{4069.583 \pm 0.008 \pm 0.049}$) in the ACS F814W image, where the first uncertainty reported is the standard deviation of the measured SN location over the 1000 trials and the second uncertainty corresponds to the median RMS uncertainty of the stars not used to calculate the astrometric transformation. 
No source is detected within five sigma of this location, as indicated in Figure~\ref{fig:snsite}.

We then processed the individual archival FLC and C0F frames through {\tt AstroDrizzle} \citep{Hack2012} to flag cosmic-ray hits, and then extracted photometry from these frames using {\tt Dolphot} \citep{Dolphin2000,Dolphin2016}. 
Given that the prognitor is not detected in the pre-explosion {\sl HST\/} images, we place upper limits on the progenitor detection which we list in Table \ref{tab:HST}.

We note in passing that, owing to the relative proximity of SN~2018ivc to the active nucleus of NGC~1068, the various pre-explosion data of the host galaxy obtained by the {\sl Spitzer Space Telescope}, even at the shortest wavelength IRAC band at 3.6 $\mu$m, are of little value for progenitor identification, since the image of the nucleus was too luminous and effectively saturated the detectors. 
Given the comparatively low {\sl Spitzer\/} spatial resolution, the pixels at the SN site were heavily affected by this saturation.

\begin{table*}[ht]
\begin{center}
\caption{{\sl HST\/} Upper Limits to the SN 2018ivc Progenitor Detection}\label{tab:HST}
\begin{tabular}{ccccccc}
\hline
Date & Instrument & Filter & Apparent Mag & Absolute Mag & Program ID & PI \\
     &            &        & Limit (3$\sigma$)       & Limit (3$\sigma$) &            &    \\
\hline
2007-08-16 & WFPC2   & F450W & $>25.8$ & $>-6.2$ & GO-11128 & D. Fisher\\
2003-01-08 & ACS/WFC & F550M & $>25.0$ & $>-6.5$ & GO-9503  & N. Nagar \\
1994-12-03 & WFPC2   & F606W & $>25.0$ & $>-6.4$ & GO-5479  & M. Malkan \\
2001-06-30 & WFPC2   & F606W & $>25.4$ & $>-6.0$ & GO-8597  & M. Regan \\
2003-10-26 & ACS/WFC & F658N & $>23.4$ & $>-7.9$ & GO-9788  & L. Ho \\
2003-10-26 & ACS/WFC & F814W & $>24.4$ & $>-6.5$ & GO-9788  & L. Ho\\
\hline
\end{tabular}
\end{center}
\end{table*}

%%%%%%%%%%%%%%%%%%%%%%%%%%%%%%%%%%%%
%--------- Discussion -------------%
%%%%%%%%%%%%%%%%%%%%%%%%%%%%%%%%%%%%

\section{Discussion}\label{sec:Discussion}
The immediate identification and high-cadence photometric and spectroscopic follow-up observations offer us a detailed picture of the evolution of SN~2018ivc that would not otherwise be possible.
Although the declining light curve indicates that this is a Type IIL-like SN, there is evidence that the progenitor is more complicated than that of the typical Type IIL-like SN.
The strong \HeI{} lines, not always visible in Type IIL-like SNe, could indicate that the progenitor lost most of its hydrogen envelope.
The rapidly declining light curve corroborates this picture of mass loss. 
The frequent change in slope suggests that, in addition to the linear decay of the small hydrogen envelope, the shock is encountering shells of different densities that were ejected from the star during its lifetime.
This would imply that some interaction between the SN and the CSM is also partially powering the light curve of SN~2018ivc, even though narrow lines typical of some interacting SNe (IIn) are not detected.

The narrow lines ($\sim 10^{2}$ \kms) of Type IIn SNe are formed by the recombination of unshocked CSM that has been ionized by photons from the forward-shock front.
CSM can also produce intermediate-width lines ($\sim 10^3$ \kms) from the recombination of gas after the shock wave has passed through it. 
In SN~2018ivc, we observe broad emission from the SN ejecta and narrow emission from the host galaxy, but find no evidence that a narrow line SN component exists (see Section \ref{sec:SpecEvolve}).
This lack of narrow lines may indicate a clumpy CSM that has been enveloped by the SN ejecta \citep{2015smith,2018andrews}.
We note, however, that Type IIn SNe do not always exhibit narrow lines at all epochs and it is possible that we do not have a high-resolution observation during the time that narrow lines were visible.
Additionally, it is possible that the narrow lines from the CSM are not strong enough to show above broad lines of the ejecta or are too narrow and weak to be distinguished from the significant host contamination.

Despite the lack of narrow lines, there are several other indications that the ejecta of SN~2018ivc are interacting with CSM.
The boxy profiles of the \Ha{} and \HeI{} $\lambda 6678$ complex could also be indicative of interaction with a shell of CSM \citep{Andrews2010,Inserra2011}. 
The strong X-ray detection likely originates from the shocked CSM. 
Together, the light curve and spectroscopic observations demonstrate the presence of interaction in SN~2018ivc.

The HV features seen in SN~2018ivc are unusual.
The presence of these features indicates that hot, dense, asymmetric material is moving with the speed of the ejecta (if the material is in the line of sight) or faster.
Given that the HV features are present in hydrogen, helium, and calcium emission, this feature may be due to material that was ejected in the explosion.
It is possible that a bullet of $\rm{{}^{56}Ni}$ was ejected early in the explosion at high speed and that its the radioactive decay powers these features. 
 
Searching the literature, we find that while multicomponent hydrogen features are seen in Type IIP/IIL SNe, they are most often during the nebular phase and at significantly lower velocity.
Two notable exceptions to this are SN~2014G \citep{2016terreran} and SN~2010jp \citep{2012smith}.
SN~2014G was a Type IIL-like SN which showed narrow flash spectroscopic features during the first 10 days of evolution. 
Around day 100, SN~2014G developed a narrow feature blueward of \Ha{} that could not be associated with any other species and thus was identified as a HV hydrogen feature, with an initial velocity of $\sim 7600$ \kms{}.
\citet{2016terreran} explain this feature as being caused by the spherically symmetric SN ejecta interacting with a bipolar lobe CSM with a $40^{\circ}$ angle between the CSM axis and the observer's line of sight.
SN~2010jp was a low luminosity, linearly declining Type IIn SN which showed a triple-peaked \Ha{} emission line.
The red and blue \Ha{} peaks ($-12,000$ and 15,000 \kms{}, respectively) are explained by a jet-powered explosion. 
The HV features in SN~2018ivc, at comparable speeds to those of SN~2014G and SN~2010jp, could originate from a disk or jet-like structure. 

\begin{figure*}[ht]
    \centering
    \includegraphics[width=\textwidth]{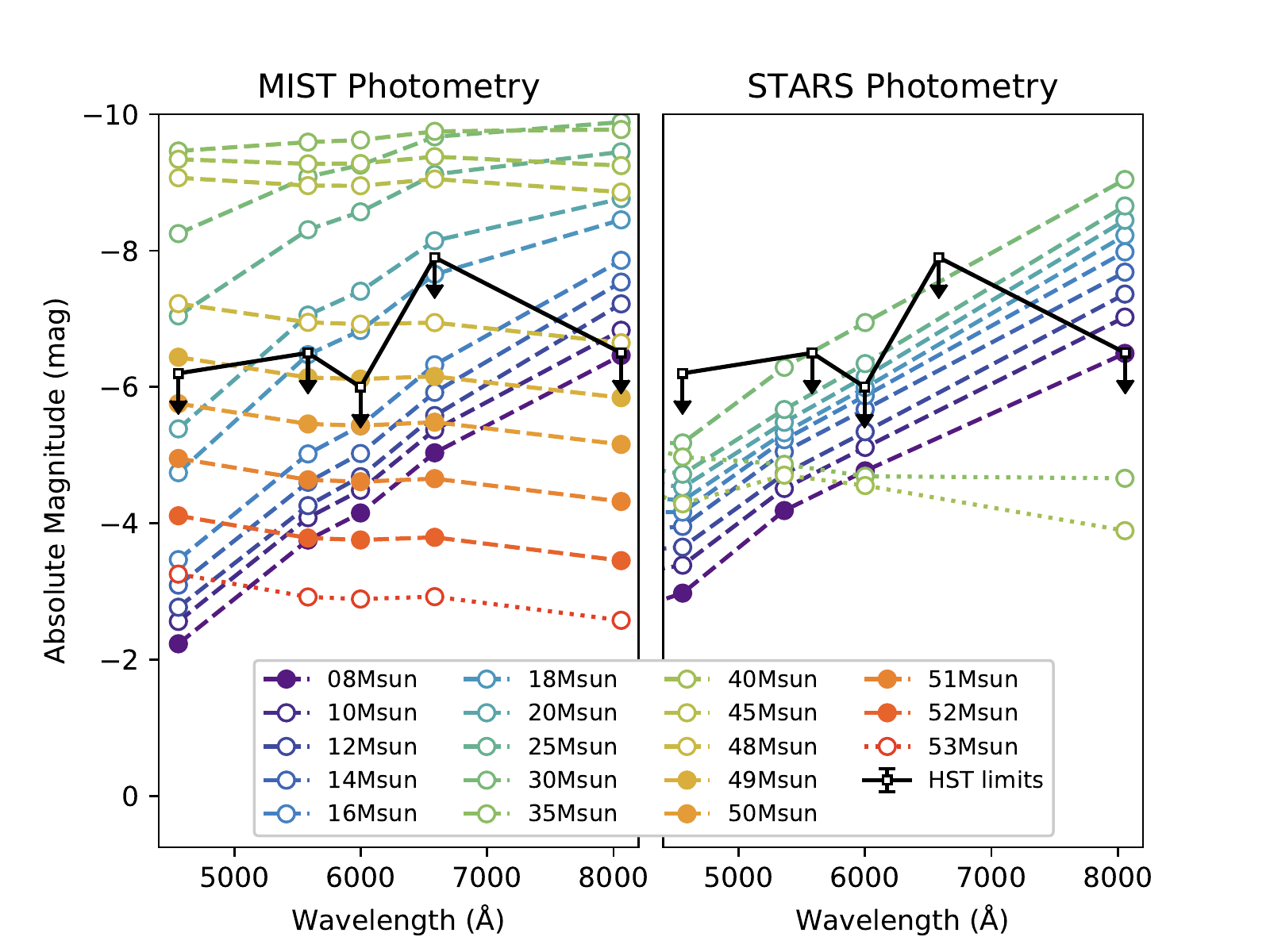}
    \caption{Upper limits on the detection, as given in Table~\ref{tab:HST}, of the SN 2018ivc progenitor in archival {\sl HST\/} images (black squares and arrows). In color are  model SEDs derived from the endpoint of the solar-metallicity MIST \citep{Choi2016,Dotter2016,Paxton2011,Paxton2013,Paxton2015} single-star evolutionary tracks (left) and the {\tt STARS\/} models (\citealt{Eggleton1971,Pols1995,Eldridge2004} (right) for masses spanning the range of allowed masses for each evolutionary model code. 
    Allowed models are denoted with filled color circles. 
    Models that are ruled out by the upper limits are marked with open circles and connected by a dashed line.
    Models that are ruled out by a lack of hydrogen in their envelope ($M\rm{_{H, env}}<0.1$ \msun{}) are denoted with open circles and connected by a dotted line.
    The {\sl HST\/} upper limits and strong hydrogen features in the spectra imply that that progenitor of SN~2018ivc, if single, was likely 8 \msun{}. 
    The MIST models allow additional progenitors in the range 49-52 \msun{}, although these have small hydrogen envelopes which may not be able to produce the features seen in the spectra of SN~2018ivc.}
    \label{fig:limits}
\end{figure*}

Based on the upper limits we derive from the {\sl HST\/} pre-explosion images (see Table~\ref{tab:HST} and Figure~\ref{fig:limits}) and the fact that hydrogen is visible and strong throughout the spectroscopic evolution, we evaluate what we can infer about the progenitor.
Referring to the single-star evolutionary tracks at solar metallicity from the MESA Isochrones \& Stellar Tracks (MIST v1.2; \citealt{Choi2016,Dotter2016,Paxton2011, Paxton2013, Paxton2015}), we exclude stars whose photometry would exceed our upper limits or with less than 0.1 \msun{} of hydrogen in the envelope. 
With these criteria we find we can eliminate all stars with initial masses between 9 \msun{} and 48 \msun{} and above 52 \msun{} as the progenitor (see left panel of Figure~\ref{fig:limits}). 
We conservatively include stars between 49 \msun{} and 52 \msun{} although we note that the largest hydrogen envelope mass in this range is 0.5 \msun{}, which could possibly be excluded with detailed spectroscopic modeling that is beyond the scope of this paper.
Additionally, we caution that this limit is highly dependent on the evolutionary models used. 
For instance, if we adopt the {\tt STARS\/} models (\citealt{Eggleton1971,Pols1995,Eldridge2004}; on which the BPASS binary evolution models are based, e.g., \citealt{Eldridge2017}), we find that the hydrogen mass in the envelope drops below 0.1\msun{} at a lower mass and redder SED, resulting in the exclusion of all stars with masses greater than 8 \msun{} (see right panel of Figure~\ref{fig:limits}).

\begin{figure}[ht]
    \centering
    \includegraphics[width=\columnwidth]{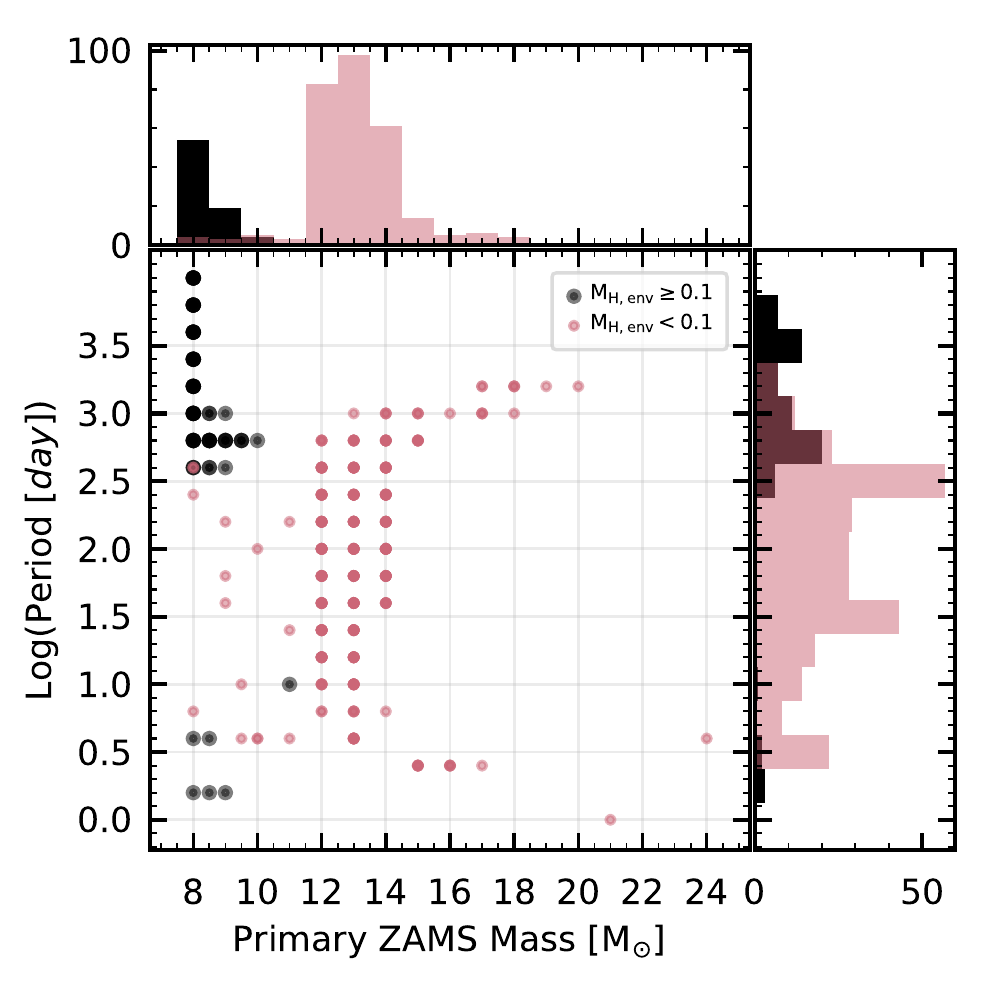}
    \caption{The range of progenitor masses and periods of the solar metallicity model binary systems from BPASS \citep{Eldridge2017} that are allowed by the pre-explosion upper limits.
    Models that also contain more than 0.1 \msun{} of hydrogen in their envelopes are marked as black circles, while models that contain less hydrogen are marked as pink points.
    The model parameters are marginalized over the ratio of the secondary to primary star mass (see Figure \ref{fig:3d} for an animation of the three-dimensional distribution).
    The opacity of the markers indicates the density of models at a given location.
    The distribution of progenitor masses, further marginalized over the period, is shown in the top panel, with allowed progenitors in black and eliminated progenitors in pink.
    Similarly, the distribution of periods, further marginalized over progenitor mass, is shown in the right panel with the same color scheme.
    Consistent with the single stars, we find that the progenitor was less than 12 \msun{} and most probably 8 \msun{}.}
    \label{fig:BinaryLim}
\end{figure}

\begin{figure}
    \begin{interactive}{animation}{BPASS_3d.mp4}
    \includegraphics[width=\columnwidth]{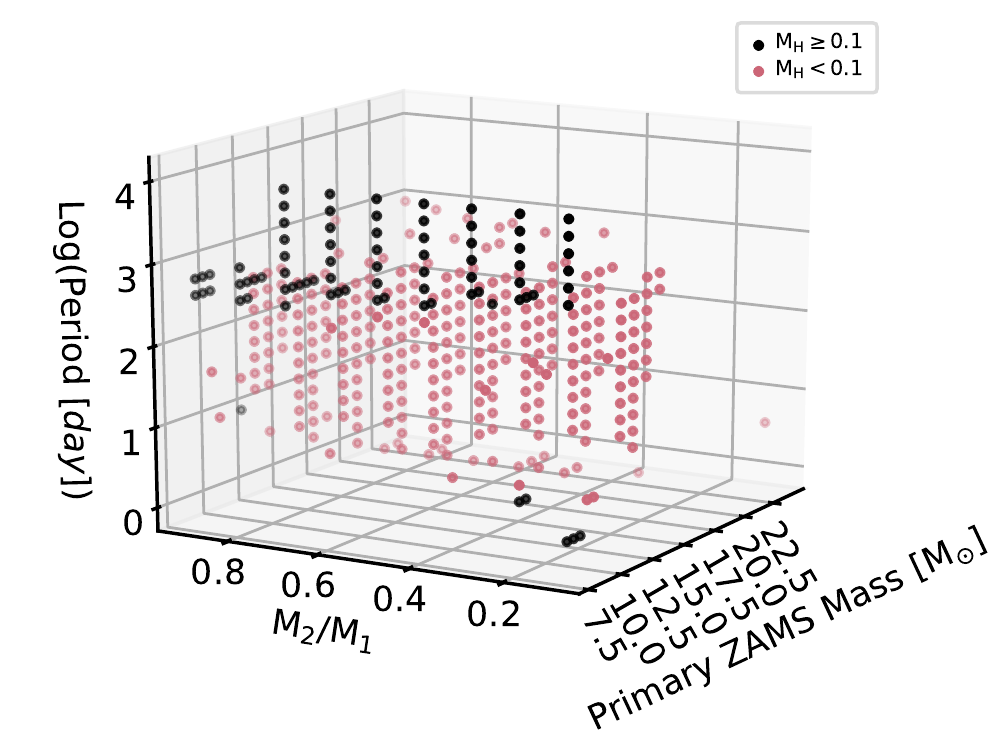}
    \end{interactive}
    \caption{An animated version of this figure is available in the online version of this paper. 
    The range of progenitor mass, period, and mass ratio of the secondary to the primary star of the solar metallicity model binary systems from BPASS \citep{Eldridge2017} that are allowed by the pre-explosion upper limits.
    Models that are allowed by both the {\sl HST\/} upper limits and the hydrogen in their envelopes are shown in black, models that are ruled out are shown in pink.
    Most mass ratios are possible for allowed $\rm{log(Period [days])}$ and progenitor masses, although there are 6 models that only occur at specific combinations of all three parameters.}
    \label{fig:3d}
\end{figure}

The majority of massive stars form in binary systems \citep{2012sana}.
For this reason, we also consider possible binary progenitor systems by examining the endpoints of the BPASS v2.2 models and the light from each combined system.
Again, considering the {\sl HST\/} upper limits and the mass of hydrogen in the envelope, we find a maximum progenitor mass of 11 \msun{}, with the majority of progenitors between 8 \msun{} and 9 \msun{} (see Figure~\ref{fig:BinaryLim}). 
Most progenitor systems are in a wide binary with $\rm{log(Period [day])=2.75}$ for masses between 8 \msun{} and 11 \msun{} and a broader range of $\rm{log(Period)}$ (2.5-4) for an 8 \msun{} progenitor.
The few short-period progenitors occur when the secondary mass is much smaller than the primary mass ($\rm{M_{2}/M_{1}\leq0.2}$; see Figure \ref{fig:3d}).

To better understand the nature of SN~2018ivc and its progenitor, we compare the host properties derived in Section \ref{sec:host} with those of all Type IIP/IIL SNe (85) and Type IIn SNe (16) from the PMAS/PPak Integral-field Supernova hosts COmpilation (PISCO) sample \citep{Galbany2018}\footnote{Updated with all new observations obtained through May 2019.}.
The metallicity of the parent \HII{} region of SN~2018ivc is near the median of the Type IIP/IIL SN distribution and slightly below the median of the Type IIn SN distribution (see the upper-right panel of Figure \ref{fig:host}).
Similarly, the EW\Ha{} falls near the median of both Type IIP/IIL SN and Type IIn SN distributions. 

EW\Ha{} is an indicator of the age of the cluster.
\citet{Kuncarayakti2013} show a theoretical relationship between EW\Ha{} and the age of the stellar population assuming a Salpeter initial mass function (IMF) and an instantaneous burst of star formation. 
The age of the stellar population can then be used to estimate the age of the SN progenitor. 
Using the EW\Ha{} derived from the parent \HII{} region and Figure 1 from \citet{Kuncarayakti2013}, we find the age of the progenitor of SN~2018ivc to be 6.75--7.75\,Myr (depending on the slope and upper value of the IMF), corresponding to a progenitor mass of 25--28\,M$_{\rm{\odot}}$.
This value is similar to the 25\,M$_{\rm{\odot}}$ of SN~1996al found by \citet{2016benetti}.
We note, however, that this agreement may be coincidental owing to limitations in our ability to isolate a single stellar population at this distance as well as simplifying assumptions made in mapping of the EW\Ha{} modeling to progenitor age (see \citealt{2019schady} for a detailed explanation). 
Specifically, the EW\Ha{}-age relation used assumes that massive stars are the ionizing source producing the \Ha{} emission, that we recover all of the photons ionizing the surrounding gas (i.e., there is no leakage), and that there are no binaries in the stellar population.
The inclusion of these effects could yield an older age and therefore a lower mass progenitor, which is more consistent with the constraints from pre-explosion imaging.

%%%%%%%%%%%%%%%%%%%%%%%%%%%%%%%%%%%%
%--------- Summary -------------%
%%%%%%%%%%%%%%%%%%%%%%%%%%%%%%%%%%%%
\section{Summary}\label{sec:conclude}
In this paper we have described the early discovery and prompt follow-up observations of SN~2018ivc by the DLT40 team and high-cadence monitoring by the GSP over the first 80 days of evolution.
The DLT40 survey observed SN~2018ivc, identified it as a SN candidate using automated software, and confirmed the SN with follow-up observations within 15\,min of the discovery observation.
This discovery triggered a comprehensive set of photometric and spectroscopic observations by the DLT40 team and the GSP for the duration of visibility, $\sim80$ days.

The light curve of this SN changed slope every $\approx10$ days for the first 30 days before settling onto a linear decline. 
The spectroscopic evolution is rapid and the spectrum is dominated by H and \HeI{} emission lines with shallow or no P Cygni absorption.
These characteristics combined with the X-ray detection suggest that the SN ejecta are interacting with CSM.
We find HV emission in hydrogen, helium, and calcium, indicating an asymmetric progenitor or explosion.
We analyze pre-explosion IFU observations of NGC~1068 and find that SN~2018ivc exploded in a region of typical metallicity and star formation with respect to other Type II SNe. 
By analyzing the EW\Ha{} of the SN's parent \HII{} region, we find evidence that the progenitor of SN~2018ivc had an initial mass of $\rm{\leq 25\ M_{\odot}}$.

We show that SN~2018ivc resembles SN~1996al both photometrically and spectroscopically, indicating that the progenitor of SN~2018ivc may have been similar to the 25\,$\rm{M_{\odot}}$ progenitor inferred for SN~1996al, although this is in tension with the progenitors derived from pre-explosion observations.
Finally, we use {\sl HST\/} archival observations of NGC~1068, taken prior to explosion and ToO {\sl HST\/} observations of the SN to derive photometric upper limits on the progenitor.
From these limits and the strong presence of hydrogen in the spectra, we infer a probable low-mass progenitor ($M<12$ \msun{} for binary models and $M<8$ \msun{} for single-star models) although the MIST models do allow for a massive progenitor ($49\leq M \leq 52$ \msun{}).

It is only with the early discovery, immediate response, and high-cadence monitoring of SNe, like those provided by the DLT40 and GSP teams, that the complex and rapid evolution of interacting SNe like SN~2018ivc can be observed and characterized.
It is possible that the interaction in other Type IIL-like SNe, lacking narrow lines and without early identification and sufficient follow-up observations, has not been identified.
For a clear and detailed understanding of the progenitor systems of Type II SNe, specifically the role of mass loss in stellar evolution and its observational signatures, we must continue to identify SNe early, announce the discovery immediately, and combine our resources for high-cadence observations of the full SN evolution.

\acknowledgments
We are grateful to B. J. Fulton for providing us the image shown in Figure~1, to P. Chandra for analysis of the \textit{Swift} XRT and GMRT data, and to Jared Goldberg, Charlie Kilpatrick, Morgan Fraser, and Stephen Smartt for insightful conversations.  
Research by K.A.B, S.V., and Y.D. is supported by NSF grant AST--1813176.   
Research by D.J.S. is supported by NSF grants AST--1821967, 1821987, 1813708, 1813466, and 1908972.
Research by J.E.A. and N.S. was supported by NSF grant AST--151559.
L.G. was funded by the European Union's Horizon 2020 research and innovation programme under the Marie Sk\l{}odowska-Curie grant agreement No. 839090.
J.B., D.H., C.M., and D.A.H. are supported by NSF grant AST--1313484.  
I.A. is a CIFAR Azrieli Global Scholar in the Gravity and the Extreme Universe Program and acknowledges support from that program, from the Israel Science Foundation (grant numbers 2108/18 and 2752/19), from the United States - Israel Binational Science Foundation (BSF), and from the Israeli Council for Higher Education Alon Fellowship.
A.V.F. is grateful for financial assistance from the Christopher R. Redlich Fund, the TABASGO Foundation, and the Miller Institute for Basic Research in Science (U.C. Berkeley).
The research of J.C.W. is supported by NSF grant AST--1813825.
K.M. acknowledges the support from Department of Science and Technology (DST), Government of India and Indo-US Science and Technology Forum (IUSSTF) for the WISTEMM fellowship and the Department of Physics, UC Davis, where part of this work was carried out.
P.J.B.'s work on SOUSA and core-collapse supernovae is supported by NASA ADAP grants NNX13AF35G and NNX17AF43G.
D.P. gratefully acknowledges support  provided by the National Aeronautics and Space Administration (NASA) through Chandra Award Number GO8-19051X issued by the Chandra X-ray Center, which is operated by the Smithsonian Astrophysical Observatory for and on behalf of NASA under contract NAS8-03060.
J.Z. is supported by the National Natural Science Foundation of China (NSFC, grants 11773067 and 11403096), the Youth Innovation Promotion Association of the CAS (grant 2018081), and  the Western Light Youth Project.
This work of X.W. is supported by the National Natural Science Foundation of China (NSFC grants 11325313 and 11633002), and the National Program on Key Research and Development Project (grant 2016YFA0400803).
E.B. and J.D. are supported in part by NASA grant NNX16AB25G.
This work was partially supported by the Open Project Program of the Key Laboratory of Optical Astronomy, National Astronomical Observatories, Chinese Academy of Sciences. 

We thank the support staffs at the many observatories where data were obtained.
%LJT
Funding for the Li-Jiang 2.4~m telescope (LJT) has been provided by Chinese Academy of Sciences and the People's Government of Yunnan Province.
The LJT is jointly operated and administrated by Yunnan Observatories and Center for Astronomical Mega-Science, CAS.
%LCO
This work makes use of observations from the Las Cumbres Observatory network.
%Gemini
Based in part on observations obtained at the Gemini Observatory under programs  GS-2018B-Q-130 (PI: D. Sand).  Gemini is operated by the Association of Universities for Research in Astronomy, Inc., under a cooperative agreement with the NSF on behalf of the Gemini partnership: the NSF (United States), the National Research Council (Canada), CONICYT (Chile), Ministerio de Ciencia, Tecnolog\'ia e Innovaci\'on Productiva (Argentina), and Minist\'erio da Ci\^encia, Tecnologia e Inova\c{c}\~ao (Brazil). The data were processed using the Gemini IRAF package. We thank the queue service observers and technical support staff at Gemini Observatory for their assistance.
%HST
This work is based in part on observations made with the NASA/ESA {\sl Hubble Space Telescope}, obtained from the Data Archive at the Space Telescope Science Institute (STScI), which is operated by the Association of Universities for Research in Astronomy (AURA), Inc., under 
NASA contract NAS 5-26555. Support for program GO-15151 was provided by NASA through a grant from STScI.
%LBT
The LBT is an international collaboration among institutions in the United States, Italy, and Germany. LBT Corporation partners are The University of Arizona on behalf of the Arizona university system; Istituto Nazionale di Astrofisica, Italy; LBT Beteiligungsgesellschaft, Germany, representing the Max-Planck Society, the Astrophysical Institute Potsdam, and Heidelberg University; The Ohio State University; and The Research Corporation, on behalf of The University of Notre Dame, University of Minnesota, and University of Virginia.
%NED
This research has made use of the NASA/IPAC Extragalactic Database (NED) which is operated by the Jet Propulsion Laboratory, California Institute of Technology, under contract with NASA.
%Stewart and MMT:
Observations using Steward Observatory facilities were obtained as part of the large observing programme AZTEC: Arizona Transient Exploration and Characterization. 
Some observations reported here were obtained at the MMT Observatory, a joint facility of the University of Arizona and the Smithsonian Institution.
%LBT
The LBT/MODS observations were obtained as part of a pilot effort at Steward Observatory to move to queue observing and include back-up programs for poor weather conditions or technical failures.
We thank J. Power for coordinating the queue efforts for our LBT/MODS observations.
%SOAR
Based in part on observations obtained at the Southern Astrophysical Research (SOAR) telescope, which is a joint project of the Minist\'{e}rio da Ci\^{e}ncia, Tecnologia, Inova\c{c}\~{o}es e Comunica\c{c}\~{o}es (MCTIC) do Brasil, the U.S. National Optical Astronomy Observatory (NOAO), the University of North Carolina at Chapel Hill (UNC), and Michigan State University (MSU).
%IIA
The facilities at IAO and CREST are operated by the Indian Institute of Astrophysics, Bangalore.
%Chandra
This research has made use of data obtained from the Chandra Data Archive and software provided by the Chandra X-ray Center (CXC) in the application packages CIAO and Sherpa.
%Keck
Some of the data presented herein were obtained at the W. M. Keck Observatory, which is operated as a scientific partnership among the California Institute of Technology, the University of California, and NASA; the observatory was made possible by the generous financial support of the W. M. Keck Foundation.

%% To help institutions obtain information on the effectiveness of their 
%% telescopes the AAS Journals has created a group of keywords for telescope 
%% facilities.
%
%% Following the acknowledgments section, use the following syntax and the
%% \facility{} or \facilities{} macros to list the keywords of facilities used 
%% in the research for the paper.  Each keyword is check against the master 
%% list during copy editing.  Individual instruments can be provided in 
%% parentheses, after the keyword, but they are not verified.

\vspace{5mm}
\facilities{ARC (DIS), ARIES:ST, ARIES:DFOT, Beijing:2.16m (BFOSC), CTIO:PROMPT, Gemini:South (Flamingos-2), CXO, HCT (HFOSC), HET (LRS2), {\sl HST} (ACS, WFPC2, WFC3), Keck: I (LRIS), II (DEIMOS), Las Cumbres Observatory (FLOYDS, Sinistro), LBT (MODS), MAST, MMT (Binospec),  PO:1.2m, SALT (RSS), SOAR (Goodman),  SO: Bok (SPOL), SO: Kuiper (Mont4K) SO:Super-LOTIS, UMN:1.52m}, {\sl Spitzer} (IRAC, MIPS), Swift (UVOT), YAO:2.4m (YFOSC)

%% Similar to \facility{}, there is the optional \software command to allow 
%% authors a place to specify which programs were used during the creation of 
%% the manusscript. Authors should list each code and include either a
%% citation or url to the code inside ()s when available.
\software{
astropy \citep{2013A&A...558A..33A,astropy},  {\sc lcogtsnpipe} \citep{Valenti_2016}, CIAO \citep{2006SPIE.6270E..1VF}, Sherpa \citep{2001SPIE.4477...76F}, SpectRes \citep{2017Carnall}}

%% Appendix material should be preceded with a single \appendix command.
%% There should be a \section command for each appendix. Mark appendix
%% subsections with the same markup you use in the main body of the paper.

%% Each Appendix (indicated with \section) will be lettered A, B, C, etc.
%% The equation counter will reset when it encounters the \appendix
%% command and will number appendix equations (A1), (A2), etc. The
%% Figure and Table counter will not reset.

\appendix
\section{Spectroscopic Observations}
\begin{longtable*}{cccccccc}
%\begin{center}
\caption{Log of Spectroscopic Observations\label{tab:spec}}\\
%\begin{tabular}{cccccccc}
\hline
\hline
Observation & MJD & Phase & Telescope & Instrument & Wavelength & Exposure & Resolution\\
Date &  & (day) & & & Range (\rm{\AA}) & Time (s) & $\rm{ \lambda/\Delta\lambda} $\\
\hline
\endhead
2018-11-24 & 58446.29 & 2.04 & FTN & FLOYDS & 3499-10000 & 2700 & 400-700 \\
2018-11-24 & 58446.50 & 2.25 & 2.16m XLO & BFOSC & 3852-8696 & 3000 & 250-800 \\
2018-11-24 & 58446.70 & 2.45 & 2.4m LJT & YFOSC & 3400-9100 & 1200 & 240 \\
2018-11-24 & 58446.81 & 2.56 & SALT & RSS & 3494-9392 & 1793 & 600-2000 \\
2018-11-25 & 58447.20 & 2.95 & Gemini-S & FLAMINGOS-2 & 9852-18075 & 720 & 900 \\
2018-11-25 & 58447.38 & 3.13 & FTN & FLOYDS & 3499-10000 & 1200 & 400-700 \\
2018-11-25 & 58447.52 & 3.27 & 2.16m XLO & BFOSC & 3852-8696 & 3300 & 250-800 \\
2018-11-25 & 58447.60 & 3.35 & HCT-IIA & HFOSC & 3500-8998 & 1200 & 1200 \\
2018-11-26 & 58448.57 & 4.32 & 2.16m XLO & BFOSC & 3852-8697 & 3300 & 250-800 \\
2018-11-27 & 58449.16 & 4.91 & HET & lrs2 & 3640-10298 & 606 & 1140-1920 \\
2018-11-27 & 58449.32 & 5.07 & LBT & MODS & 3290-5549 & 2400 & 1850-2300 \\
2018-11-27 & 58449.32 & 5.07 & LBT & MODS & 5800-9572 & 2400 & 1850-2300 \\
2018-11-27 & 58449.61 & 5.36 & 2.16m XLO & BFOSC & 3853-8701 & 3300 & 250-800 \\
2018-11-28 & 58450.23 & 5.98 & FTN & FLOYDS & 3499-10000 & 1200 & 400-700 \\
2018-11-29 & 58451.25 & 7.00 & Bok & SPOL & 4001-7549 & 2400 & 430 \\
2018-12-01 & 58453.40 & 9.15 & FTN & FLOYDS & 3499-10000 & 1200 & 400-700 \\
2018-12-02 & 58454.00 & 9.75 & Gemini-S & FLAMINGOS-2 & 9851-18077 & 1080 & 900 \\
2018-12-03 & 58455.69 & 11.44 & HCT-IIA & HFOSC & 3500-8998 & 1200 & 1200 \\
2018-12-04 & 58456.59 & 12.34 & HCT-IIA & HFOSC & 3500-8998 & 1200 & 1200 \\
2018-12-05 & 58457.44 & 13.19 & Keck I & LRIS & 3136-10220 & 119 & 750-1475 \\
2018-12-07 & 58459.20 & 14.95 & Gemini-S & FLAMINGOS-2 & 9849-18077 & 1080 & 900 \\
2018-12-07 & 58459.55 & 15.30 & HCT-IIA & HFOSC & 3500-8998 & 1800 & 1200 \\
2018-12-08 & 58460.36 & 16.11 & FTN & FLOYDS & 3500-9999 & 1200 & 400-700 \\
2018-12-10 & 58462.32 & 18.07 & 3.5m APO & DIS & 3374-5607 & 2400 & 2450 \\
2018-12-10 & 58462.32 & 18.07 & 3.5m APO & DIS & 5263-9404 & 2400 & 3150 \\
2018-12-11 & 58463.81 & 19.56 & HCT-IIA & HFOSC & 3500-8998 & 2400 & 1200 \\
2018-12-13 & 58465.00 & 20.75 & Gemini-S & FLAMINGOS-2 & 9850-18077 & 1440 & 900 \\
2018-12-14 & 58466.30 & 22.05 & MMT & Binospec & 5072-7541 & 720 & 3590 \\
2018-12-15 & 58467.76 & 23.51 & HCT-IIA & HFOSC & 3500-8998 & 2400 & 1200 \\
2018-12-16 & 58468.65 & 24.40 & HCT-IIA & HFOSC & 3500-8998 & 2400 & 1200 \\
2018-12-17 & 58469.23 & 24.98 & FTN & FLOYDS & 3500-10000 & 1800 & 400-700 \\
2018-12-18 & 58470.51 & 26.26 & 2.16m XLO & BFOSC & 4358-8690 & 2400 & 250-800 \\
2018-12-19 & 58471.10 & 26.85 & Gemini-S & FLAMINGOS-2 & 9852-18076 & 1800 & 900 \\
2018-12-21 & 58473.15 & 28.90 & SOAR & GHTS & 3500-7000 & 780 & 1050 \\
2018-12-21 & 58473.15 & 28.90 & SOAR & GHTS & 5000-9000 & 780 & 1050 \\
2018-12-22 & 58474.35 & 30.10 & FTN & FLOYDS & 3499-9999 & 3600 & 400-700 \\
2018-12-23 & 58475.21 & 30.96 & FTN & FLOYDS & 3499-9999 & 3600 & 400-700 \\
2018-12-24 & 58476.63 & 32.38 & HCT-IIA & HFOSC & 3500-8998 & 2400 & 1200 \\
2018-12-24 & 58476.78 & 32.53 & HCT-IIA & HFOSC & 3500-8998 & 1200 & 1200 \\
2018-12-24 & 58476.85 & 32.60 & SALT & RSS & 6057-7010 & 2464 & 2200-5500 \\
2018-12-26 & 58478.10 & 33.85 & Gemini-S & FLAMINGOS-2 & 9849-18077 & 1800 & 900 \\
2018-12-26 & 58478.48 & 34.23 & FTS & FLOYDS & 4800-10000 & 3600 & 400-700 \\
2018-12-26 & 58478.83 & 34.58 & SALT & RSS & 3497-9393 & 2093 & 600-2000 \\
2018-12-30 & 58482.49 & 38.24 & FTS & FLOYDS & 4800-9999 & 3600 & 400-700 \\
2019-01-03 & 58486.30 & 42.05 & FTN & FLOYDS & 3500-10000 & 3600 & 400-700 \\
2019-01-05 & 58488.00 & 43.75 & Gemini-S & FLAMINGOS-2 & 9851-18075 & 1800 & 900 \\
2019-01-06 & 58489.54 & 45.29 & 2.16m XLO & BFOSC & 3851-8692 & 3600 & 250-800 \\
2019-01-10 & 58493.33 & 49.08 & FTN & FLOYDS & 3499-10000 & 3600 & 400-700 \\
2019-01-11 & 58494.30 & 50.05 & Keck I & LRIS & 3137-10197 & 420 & 750-1475 \\
2019-01-13 & 58496.49 & 52.24 & 2.16m XLO & BFOSC & 4095-8810 & 3000 & 250-800 \\
2019-01-14 & 58497.10 & 52.85 & Gemini-S & FLAMINGOS-2 & 9850-18079 & 2160 & 900 \\
2019-01-22 & 58505.26 & 61.01 & FTN & FLOYDS & 3499-9999 & 3600 & 400-700 \\
2019-01-28 & 58511.44 & 67.19 & 2.16m XLO & BFOSC & 3869-8822 & 3600 & 250-800 \\
2019-02-05 & 58519.25 & 75.00 & Keck I & LRIS & 3138-10244 & 1200 & 750-1475 \\
2019-02-08 & 58522.13 & 77.88 & MMT & Binospec & 5062-7522 & 960 & 3590 \\
2019-08-28 & 58723.59 & 279.34 & Keck II & DEIMOS & 4480-9510 & 1200 & 1875\\
\hline
%\end{tabular}
%\end{center}
\end{longtable*}

%% The reference list follows the main body and any appendices.
%% Use LaTeX's thebibliography environment to mark up your reference list.
%% Note \begin{thebibliography} is followed by an empty set of
%% curly braces.  If you forget this, LaTeX will generate the error
%% "Perhaps a missing \item?".
%%
%% thebibliography produces citations in the text using \bibitem-\cite
%% cross-referencing. Each reference is preceded by a
%% \bibitem command that defines in curly braces the KEY that corresponds
%% to the KEY in the \cite commands (see the first section above).
%% Make sure that you provide a unique KEY for every \bibitem or else the
%% paper will not LaTeX. The square brackets should contain
%% the citation text that LaTeX will insert in
%% place of the \cite commands.

%% We have used macros to produce journal name abbreviations.
%% \aastex provides a number of these for the more frequently-cited journals.
%% See the Author Guide for a list of them.

%% Note that the style of the \bibitem labels (in []) is slightly
%% different from previous examples.  The natbib system solves a host
%% of citation expression problems, but it is necessary to clearly
%% delimit the year from the author name used in the citation.
%% See the natbib documentation for more details and options.

\bibliographystyle{aasjournal}
\bibliography{biblio}

%% This command is needed to show the entire author+affilation list when
%% the collaboration and author truncation commands are used.  It has to
%% go at the end of the manuscript.
%\allauthors

%% Include this line if you are using the \added, \replaced, \deleted
%% commands to see a summary list of all changes at the end of the article.
%\listofchanges

\end{document}